\documentclass[12pt]{iopart}
\usepackage{iopams}
\usepackage{cleveref}
\usepackage{breqn}
\usepackage{calligra}
\usepackage{array,multirow}
\usepackage[table]{xcolor}
\usepackage{graphicx} 
\usepackage{cite}
\usepackage{amsopn}

\newcolumntype{L}[1]{>{\raggedright\let\newline\\\arraybackslash\hspace{0pt}}m{#1}}
\newcolumntype{C}[1]{>{\centering\let\newline\\\arraybackslash\hspace{0pt}}m{#1}}
\newcolumntype{R}[1]{>{\raggedleft\let\newline\\\arraybackslash\hspace{0pt}}m{#1}}

\newcommand\ajustspaceandequationnumber{
   \vspace{-1. \belowdisplayskip}
   \vspace{-1.05\abovedisplayskip}
      }

\newcommand{\operator}[1]{\widehat{#1}}

\DeclareMathOperator{\pcheck}{\operator{H}}

\makeatletter
\let\cref@old@eq@setnumberOld\eq@setnumber
\def\eq@setnumber{%
\cref@old@eq@setnumberOld%
\cref@constructprefix{equation}{\cref@result}%
\protected@xdef\cref@currentlabel{%
[equation][\arabic{equation}][\cref@result]\p@equation\eq@number}}
\makeatother

\crefname{equation}{Eq.}{Eqs.}
\crefname{table}{Tab.}{Tabs.}
\crefname{figure}{Fig.}{Figs.}
\crefname{appsec}{Appendix}{Appendices}

\crefrangeformat{equation}{Eqs.~(#3#1#4--#5#2#6)}
\crefmultiformat{equation}{Eqs.~(#2#1#3}{--#2#1#3)}{}{--#2#1#3)}
\Crefmultiformat{equation}{Eqs.~(#2#1#3}{--#2#1#3)}{}{--#2#1#3)}
\crefrangelabelformat{equation}{#3#1#4--#5\crefstripprefix{#1}{#2}#6}

\crefformat{table}{Tab.~(#2#1#3)}
\crefformat{figure}{Fig.~(#2#1#3)}

\newcommand{\sect}[1]{Sec. \cref{#1}}

\newcommand{\co}{k} 
 
\newcommand{\dm}{\boldsymbol{d}} 
\newcommand{\dmc}[1]{d_{#1}}
\newcommand{\dmz}{\boldsymbol{0}} 
\newcommand{\ndm}{n} 
\newcommand{\um}[1]{\boldsymbol{u_{#1}}}
\newcommand{\Ndm}{m_\mathrm{d}} 
\newcommand{\Nem}{m_\mathrm{e}} 
\newcommand{\Nndm}{m_\mathrm{n}} 
\newcommand{\Nmm}{m_\mathrm{m}} 
\newcommand{\Mdm}{\overline{m_\mathrm{d}}} 
 
\newcommand{\Mndm}{\overline{m_\mathrm{n}}} 
\newcommand{\Mmm}{\overline{m_\mathrm{m}}}  
\newcommand{\Mco}{\overline{\co}}  
\newcommand{\stt}{\sigma=\{\dm,\Nndm,\Nmm,\co\}} 
\newcommand{\sttz}{\sigma_0=\{\dmz,0,0,\co\}}

\newcommand{\de}{{}\,\mathrm{d}}

\newcommand{\laplace}{\mathcal{L}}

\newcommand{\rdm}{r_\mathrm{d}}

\newcommand{\rndm}{r_\mathrm{n}}

\newcommand{\rap}{r_\mathrm{apop}}
\newcommand{\ras}{r_{\mathrm{asym}}}
\newcommand{\rsy}{r_\mathrm{sym}}
\newcommand{\rpa}{r_\mathrm{pass}}

\newcommand{\drug}{\mathrm{drug}}

\newcommand{\cs}{\mathit{s}} 
\newcommand{\act}{\mathrm{act}}
\newcommand{\tgfb}{\textrm{TGF--$\beta$}}

\newcount\repstringloopcounter

\newcommand{\rep}[2]{%
  \repstringloopcounter0
  \loop\ifnum\repstringloopcounter < #1 
  #2%
  \advance\repstringloopcounter by 1 %
  \repeat 
}

\newcounter{reploopcounter}

\def\repeatstring#1#2{%
  \setcounter{reploopcounter}{#1}%
  \ifnum\value{reploopcounter}>0\relax#2%
    \addtocounter{reploopcounter}{-1}%
    \def\tmp{\repeatstring{\value{reploopcounter}}{#2}}%
    \expandafter\tmp%
  \fi%
}

\newcommand{\matr}[2]
{
\def\temp{\begin{array}}%
\def\tempp{{#1}}
\left( 
\expandafter\temp\tempp
#2
\end{array}
\right) 
}

\newcommand{\matx}[1]{ \left( \begin{array}{*{10} c}  #1  \end{array} \right) }

\newcommand{\statevec}[1]{\bra{#1}}

\newcommand{\operate}[1]{\left( \bstatevec{#1} \right)}
\renewcommand{\operate}[1]{#1}

\newcommand{\inner}[2]{\left\langle \statevec{#2},#1 \right\rangle}
\renewcommand{\inner}[2]{\ket{#2}#1}

\newcommand{\mutM}{D}
\newcommand{\maskM}{C}
\newcommand{\mutV}{d}

\newcommand{\mutjevec}{\statevec{\mutV}_j}
\newcommand{\mutevec}{\statevec{\mutV}}
\newcommand{\mutop}{\operator{\mutM}}
\newcommand{\maskop}{\operator{\maskM}_\co}
\newcommand{\mutjop}{\operator{\mutM}_j}
\newcommand{\maskjop}{\operator{\maskM}_{\co j}}
\newcommand{\eval}[1]{\lambda_{#1}}
\newcommand{\muteval}{\eval{\mutM}}
\newcommand{\maskeval}{\eval{\maskM_\co}}
\newcommand{\mutjeval}{\eval{\mutV_j}}
\newcommand{\maskjeval}{\eval{\maskM_{\co j}}}
\newcommand{\id}{\operator{\mathbb{I}}}

\newcommand{\grille}{C}

\begin{document}

\title[Modelling the order of driver mutations and metabolic mutations]{Modelling the order of driver mutations and metabolic mutations as structures in cancer dynamics}

\author{Gianluca Ascolani$^1$ and Pietro Li\'o$^1$}

\address{$^1$ Computer Laboratory, University of Cambridge, UK}
\ead{\mailto{gianluca.ascolani@cl.cam.ac.uk} and \mailto{pietro.lio@cl.cam.ac.uk}}
\vspace{10pt}

\begin{abstract}
Recent works have stressed the important role that random mutations have in the development of cancer phenotype. We challenge this current view by means of bioinformatic data analysis and computational modelling approaches. Not all the mutations are equally important for the development of metastasis. The survival of cancer cells from the primary tumour site to the secondary seeding sites depends on the occurrence of very few driver mutations promoting oncogenic cell behaviours and on the order with which these mutations occur. We introduce a model in the framework of Cellular Automata to investigate the effects of metabolic mutations and mutation order on cancer stemness and tumour cell migration in bone metastasised breast cancers. 
The metabolism of the cancer cell is a key factor in its proliferation rate. Bioinformatics analysis on a cancer mutation database shows that metabolism-modifying alterations constitute an important class of key cancer mutations. Our approach models three types of mutations: drivers, 
the order of which is relevant for the dynamics,
metabolic which support cancer growth and are estimated from existing databases, and non--driver mutations. Our results provide a quantitative basis of how the order of driver mutations and the metabolic mutations in different cancer clones could impact proliferation of therapy-resistant clonal populations and patient survival. 
Further mathematical modelling of the order of mutations is presented in terms of operators.  
We believe our work is novel because it quantifies two important factors in cancer spreading models: the order of driver mutations and the effects of metabolic mutations.

\end{abstract}
\maketitle

\section{Introduction}

The development of tumours and the formation of metastasis are caused by cells which due to mutations are incapable of correctly sensing the external signalling of the environment and the surrounding cells so to
adjust their metabolic and mitotic cycle activities. The purpose of exchange of signals between the cell and the environment is limiting their proliferation which, in the case of development of the cancer disease, becomes uncontrolled. There are different ways the single cells and organisms can defend against the development of tumour, for example, DNA self-repair mechanisms, signals inducing apoptosis in case of mechanical and biochemical instabilities of a cell and a prompt immune response which is capable of recognizing cancer cells. Mutations are accumulated by cells during their life and passed as inheritance along different progenies. In absence of external causes of mutations such as the exposition to radiations or chemicals, errors in the DNA replication and crossing over during the mitotic cycle are the major sources of mutations. 
Nowak proposed a model of cancer originating from propagation and accumulation of mutations occurring since early age \cite{Nowak1}. 
Nevertheless, cancer is an ageing related disease as its development before the age of 30 years is rare. 

Mutations may generate disruptions of cell cycle proliferation checkpoints, generate cytoskeleton instabilities inducing cell death or metabolic pathway deregulations increasing the cell proliferation, variation in the antigens and receptors on the membrane which transduce external signals (for example avoiding the immune response). These mechanisms produce the spreading of the sub--populations of cells (clones) with accumulated mutations.
The populations of cancer cells in the primary and secondary sites are characterized by a large number of genes with altered gene expression due to mutations. The tumour heterogeneity results in genotypic and phenotypic (for example drug resistance) variance in the neoplastic cell populations.
The heterogeneity increases during time as a consequence of new mutations acquired by cells during subsequent proliferations. Comparing gene expressions among cells in the primary site with those in the metastatic sites, one can observe the heterogeneities between different sites are not identical in variance, as well as, in the type of mutations accumulated; furthermore, looking at overlapping subsets of mutations between different tumour sites, it is possible to derive the time a secondary site originated, and if it is due to an offspring from the primary site or a nearby secondary metastasis.      
If we consider cancer as an evolving disease in competition for oxygen and energetic resources with other healthy  cells, mutational heterogeneity represents a strategy initiated by an early small number of cancer stem cells. Therefore, the increase of heterogeneity can be seen as a diffusion process with a drift in the multidimensional space of the cell state belonging to $\{0,1\}^N$ where $N$ is the dimension of the space and represents the maximum number of genes affected by the disease during all its evolution. We believe that in order to relate cancer evolution with patient's survival we need to take into account the characteristics of cancer stem cells, the classes of mutations and for some classes, also the order of mutations.

The work is structured in the following way.
In the next subsections, we discuss the role of cancer stemness, and we define the type of mutations modelled and their effects on cells.
In the model section, we introduce the concept of order of driver mutations, and we present the corresponding mathematical formulation. After which, we describe the set of rules driving the model dynamics from which we derive the master equations in the physical time. We model the effects of metabolic mutations on the cell cycle in terms of waiting time distributions and compute the final form of the master equation depending on the transition rates. The definition of  the functional form of the transition rates in terms of the cancer stemness follows. Further discussion on the order of mutations in terms of ladder operators and the mathematical derivation of the effective driver mutations is addressed in the last method subsection.
In the results section, we present how simulations are carried out and the analysis of data supporting both the metabolic and driver mutations followed by the discussion and comparison of the three cases of interest numerically simulated.

\subsection{The role of Cancer Stemness}\label{stemness}

Stem cells are capable of both self-renewing and differentiating \cite{laplane}; this means they 
preserve
themselves during proliferation without 
undergoing 
extinction due to differentiation, and they are a source for more committed cells \cite{stemdediffer}.
The process of cell differentiation is mainly caused by epigenetic changes, and it results 
in the appearance of new cell phenotypes. These changes in the cell state are induced by external signalling or by internal variations of the cell dynamics like methylation or segregation of factors during  mitosis. 
Not all the signals and changes involved in the differentiations are persistent or permanent. 
The loss of the new acquired phenotype is called de-differentiation. 
Nevertheless, the restoration of the external niche preserving the stemness  or the circulation of factors inducing the cell stem state  might not suffice to re-establish the stem condition in differentiated cells or in cells proliferating in a stem-like favourable condition \cite{Plaks}. Therefore, differentiated cells tendentiously do not de-differentiate.

The renewal condition is met when a cell will always undergo asymmetric division or undifferentiated symmetric proliferation. Stem cells are considered renewal at the level of single cells, meaning  after proliferating a stem cell produces at least a daughter equal to itself. On the other hand, the hypotheses advanced by Knoblich and Morrison in \cite{Knoblich,Morrison} imply that stem cells might symmetrically proliferate differentiated cells when the pool of (neighbour) stem cells do not risk extinction or has reached the maximum number the niche can sustain in a stem state.
This last statement makes the definition of stem cell very close to the concept of population of cells which is in a state of high fitness given that such state is the one with more undifferentiated characteristics when compared with its progenies.    
The stemness of a cell is a behaviour which is not only present or absent, but there are different degrees of stemness actuation. 
A cell can lose its stemness when it cannot produce a more differentiated progeny or it cannot indefinitely proliferate; vice versa, a cell can increase its stemness by de-differentiation.   
 
In an elevated mutational rate condition, changes in the DNA interfere or compete with the cell differentiation process, and it may become the predominant source of stemness variations \cite{stemdediffer}. Cancer stem cells are capable of producing a heterogeneous population which presents different phenotypes. The mutations will change the internal state of the cell affecting the transcription and translation processes. Cancer cells which will become more unstable and less fit will disappear. Cancer cells with a high fitness phenotype will survive if they proliferate asymmetrically or they do not acquire too many degrading mutations. 
Cancer stemness is the behaviour of cancer cells to preserve their phenotype and generate a heterogeneous population, part of which can show more stem-like behaviours \cite{Singh}.

\subsection{Classes of mutations}

Cancer mutations have been distinguished in drivers and non--drivers (or passengers). The accumulation of evidences of clonal heterogeneity and the observation of the arise of drug resistance in clonal sub--populations, suggest that mutations usually classified as non--drivers may have an important role in the fitness of the cancer cell and in the evolution and physiopathology of cancer. Similarly, mutations that alter the metabolism may modify the fitness of the cancer cell. The evolution of cancer clones in the context of the driver and non--driver mutations has been visualized as a braided river, with the capacity to diverge and converge \cite{rivermodel}. 

Recently 122 potential immune response drivers--genetic regions in which mutations correlate with the presence or absence of immune cells infiltrating the tumours, have been reported. Current research into how tumours hide focuses on the over-expression of so-called checkpoint inhibitors in cancer cells \cite{portapardo1, portapardo2}.

There are multiple levels of heterogeneity: inter-patient, intra-patient and intra-tumour. Inter-patient heterogeneity exists at the level where even within cancer of the same type, patients exhibit differences in terms of both biology and prognosis.
In breast cancer, based on gene expression, patients can be classified into at least four broad subtypes, namely basal (triple negative breast cancer (TNBC)), HER2+, and luminal A and B subtypes. Luminal A and B are estrogen positive cancers, with luminal A having the best prognosis; HER2+ has overexpression of the HER2 growth-enhancing gene. These are slow growing tumours that respond well to treatment. The Basal or TNBC subtype is triple negative for estrogen, progestin receptors and HER2. This is the most aggressive subtype and is unresponsive to treatment. Extending on this, intra-patient heterogeneity is defined by the differences in the primary and the metastatic tumour sites; these include morphological and genetic differences, and differences in terms of tumour aggressiveness and proliferation. Intratumour heterogeneity is then explained by differences between regions within the same tumour mass \cite{ Burrell, Chan, Gerlinger1, Gerlinger2, McGranahan, Lawrence, Yap}.

The probing of the mutational space represents an advantage that cancer cell clonal populations have adopted. Also, the temporal aspect of mutations is of great importance in the evolution of the disease \cite{order2, order1}. 

In this work, we group the mutations in three main categories: driver mutations, metabolic mutations and non--driver mutations.

The driver mutations are mutations of specific genes responsible for the deregulation of pathways involved in the production of cytomembrane proteins which induce new pro-oncogenic behaviours; they increase the capability of cancer cells to self-renewal. Therefore, the driver mutations increase the fitness of cancer cells in a tissue and their capability of creating a progeny carrying all the driver and non--driver mutations accumulated. Hence, according to the definition in \cite{cancstem}, driver mutations increase the cancer cell stemness, even though it is not possible to explicitly say a cell with a higher number of driver mutations corresponds to less uncommitted cell phenotype as in the case of healthy stem cells. A relevant aspect in the evolution of the disease is that the most important driver mutationss are very few. Works from A. Trumpp \cite{baccelli} show a number between 4 and 5 drivers mutations may be responsible for high correlation with survival. Recent works have provided evidence, at least in one cancer type, the order of the driver mutations affects the cancer clone trajectories and the patient survival \cite{order1,order2}.

The role of non--driver mutations in cancer development and evolution has been modelled in \cite{mcfarland}. The authors consider two cases: 1) the neutral non--driver mutations which do not affect the survival of cells 
and 2) the deleterious non--driver mutations as opposing to the proliferative effects resulting from the pro-oncogenic driver mutations, and consequently, causing a possible melting down of the cancer cell sub--populations.  
The non--driver mutations are those mutations which do not promote the formation of cancer stem cells and do not provide new oncogenic behaviours for the cells. Non--driver mutations generally affect the genes which are not related to the pathways promoting proliferation, migration, resistance to immune response and the formation of colonies. Cancer cells acquiring non--driver mutations become more unstable both in the cytoskeletric structure and in their fitness. 

The metabolic mutations include all the set of mutations relative to the cell energetic needs and the controls of cell proliferation \cite{metabol}. We consider that cells with accumulated metabolic mutations have an accelerated proliferation cycle because there is a sustained availability of ATP, a reduced sensitivity to cell activity inhibiting signals, and many of the checking procedures which are present during the various proliferation phases are not activated. Consequently, a higher number of these mutations further speed up the cell proliferation by reducing the time necessary to activate all these processes and the costs in terms of resources and energy.

\section{Model}

\subsection{Modelling the order of driver mutations}

Although the effects of mutations on cancer heterogeneity and phenotypic plasticity of cancer evolution has led to interesting pure theoretical models (see for instance \cite{Shirayeh}), and it has also been addressed by combining theoretical modelling, bioinformatics and experimental data (see for instance \cite{sottoriva2, sottoriva1,Casasent, ductalmut} among others), there is a paucity of models that take into account the effects of the order of mutations. 
In \cite{asco2}, mutations are accumulated in a predefined sequence, and all the cells, sooner or later, will follow the same trajectory if they are not eliminated, or do not die along the way. In other cases, the gene expression of a sample of cancer cells are analysed at different times \cite{beerenwinkelphylo, elbirbo}, or in various sites \cite{Naxerova2015} in order to identify the order in which the mutations occurred \cite{Reiter,turajlic,Schwartz}. The fate of a cell is strongly related to the environment it is surrounded by. In our case, we consider that the order of mutations is given only by the environment, and it is not related to the dynamic, or evolution of the cancer cells. Indeed, the mutation order does not represent a tight  preferential order of occurrence of mutations. On the contrary, the mutation order does not induce any constraint or bias on the occurrence of mutations which is instead random. The driver mutations affect the fitness of a tumoural mutated cell in a given compartment. 
In this work, the order among the driver mutations is given by the maximization of the survival of cancer cells which travel through a sequence of compartments from primary tumour site to secondary cancer site. 
We define a path as the sequence of tissues a hypothetical cancer cell will visit during its migrations before forming a metastasis. Different cancer cells can traverse different organs or tissues defining different paths. All the paths begin at the primary tumour site, and each of them ends in a site where cancer cells can form a secondary colony. Nevertheless, due to multiple metastasization sites, the final point of the paths can differ. Generally, a path has a different probability of being traversed than others. This means that among the total amount of cells forming the primary tumour only a portion will travel through a specific path, and some of the paths will be more common than others. 

Let us define the set of $m$ compartments $\mathcal{C}=\{c_1,c_2,\ldots, c_m\}$ 
where, for convenience, $c_1$ represents the primary site and the last $s$ compartments 
$\{c_{m-s+1},\ldots,c_{m}\}$ are the secondary sites. 
Let us also introduce the transition probability $T_{ij}$ from the compartment $c_i$ to another compartment $c_j$ for those cells which already have all possible useful mutations to pass through all the compartments. The transition matrix $T$ is such that $\sum_j T_{i j}=1$, and all the elements on the diagonal are one for all the secondary sites and zero otherwise.
For each compartment $c_j$, we define the cell density $\rho_j(n)$ after $n$ transitions.
If $\sigma(i)$ is a permutation of the $m$ compartments $\mathcal{C}$ visited by the cells, then we say a path of length $l$ is given by
$$\mathcal{P}_{\mathcal{C}}(\sigma)=\{\sigma(1),\sigma(2),\ldots,\sigma(l)\}$$ 
and,  after $n\leq l$ transitions, the population which has followed the $\mathcal{P}_{\mathcal{C}}(\sigma)$ path is:
$$\rho_{\sigma(n)}=\prod_{i=1}^{n-1}T_{\sigma(i),\sigma(i+1)}\rho_{\sigma(1)}.$$
If at the beginning all the cells are in $c_1$ so that $\rho_j(n=0)=\delta_{j,1}$, then we must impose that a path' s initial point is $\sigma(1)=1$. Furthermore, a path ends when it reaches for the first time one of the absorption points meaning that, after $l$ steps, $c_{\sigma(l)}$ corresponds to one of the  secondary sites ($\sigma(l)>m-s$). Many paths will have a final population density $\rho(l)=0$ because the transition $T_{ij}=0$ before reaching the end of the path. If $\mathcal{P}_{\mathcal{C}}$ is the set of paths obtained by all possible permutations of the compartments, then $\overline{\mathcal{P}_{\mathcal{C}}}=\max_{\rho(l)}(\mathcal{P}_{\mathcal{C}})$ are the most probable paths. 
For the sake of simplicity, from here on, we restrict ourselves to the case $\overline{\mathcal{P}_{\mathcal{C}}}$ contains only one element representing the path with higher probability of being traversed. Nevertheless, there are no extra difficulties in considering multiple paths
one at a time, and then studying the jointly resultant dynamics.  
Hence, the order we consider is given by the order of the tissues (compartments) the cancer cell will visit during its migration from the primary tumour site towards the secondary metastatic sites.
The reason for introducing this type of order is related to the fact that a cancer cell migrating into a different compartment is a rare event which is almost impossible if the cell does not develop the right behaviours due to the occurrence of proper mutations.
 
A similar explanation is valid in the case of the survival (fitness landscape) of a cell while it remains for a period of time in a compartment. In different tissues, the cells need different membrane proteins to separate from the other surrounding cells, to migrate, to access or exit a specific compartment and to be recognized by the immune system as self. All these behaviours are not necessary at the same time, and in all the compartments, but only a few of them at a time are required during the travel through a compartment or in the transitions between compartments. 
As a consequence of the concomitant facts that a tumour cell needs specific driver mutations and that mutations can not be counterbalanced, the originating site and preferential way for a cell to reach the secondary sites determine the right order of accumulation of driver mutations. 
Therefore, if we order the compartments in the sequential order the cancer cells will traverse them, then we can establish the proper order of driver mutations for a high fitness and a short permanence during the development of metastases. 

After we have specified the path $\mathcal{P}$, let us introduce the index $\co\in \mathbb{N}$  identifying the compartments in the order given by the path. 
In the case the cells need only one driver mutation during each migratory transition between two compartments the index $\co$ can also be associated to the required driver mutations necessary to continue the travel along the path. If $d_\co$ is the driver mutation needed to go from compartment $\co$ to compartment $\co+1$, then the cell which follows the specific path defined as an increasing value of $\co$ at each migratory transition till compartment $\Mco=l$, will also generate the sequence  $D=\{d_1,d_2,\ldots,d_l\}$ which defines the right mutation order.

We can extend the previous case to situations where the number of driver mutations for each compartment is more than one. Hence, if $\mathcal{C}=\{c_1,\ldots,c_l\}$ is the sequence of compartment in the order visited by the cancer cells, and,  for any $\co$, $S_\co$ is the set of driver mutations necessary in the compartment $c_\co$ to improve the fitness landscape or to increase the probability of migrating in a different compartment, then we can define two sequences of sets as follow:
\begin{eqnarray}
S^\prime_\co &=\bigcup_{j=1}^{\co} S_j, \qquad\quad& 1\leq \co\leq l\label{primed}
\\
S^{\prime\prime}_\co&=S^\prime_\co\setminus S_{\co-1}, \qquad\quad& 1\leq \co\leq l \label{2primed}\\
S^{\prime\prime}_0&=\emptyset.
\end{eqnarray}  
The $\co$-th element in the sequence $\mathcal{S^{\prime}}=\{S^{\prime}_1,\ldots,S^{\prime}_l\}$ in \eref{primed} is the cumulated mutations of cells following the right order of mutations which have just entered in the $c_{\co+1}$ compartment, while the $\co$-th element in the sequence $\mathcal{S^{\prime\prime}}=\{S^{\prime\prime}_1,\ldots,S^{\prime\prime}_l\}$ in \eref{2primed} describes all the mutations for a cell following the right order of mutations that has just entered in the $c_{\co}$ compartment necessary in order to pass in the $\co+1$-th compartment.
If $S_i\cap S_j=\emptyset$ for any $1\leq i,j\leq l$, then $S^{\prime\prime}_\co=S_\co$, but in general mutations in a set $\mathcal{S}=\{S_1,\ldots,S_l\}$ can be necessary in other compartments. 
The right order of mutations, hence, also represents the most fit and the fastest path in the mutation space. Any deviation from the right order of mutations reduces to an increased cancer cells apoptosis rate and a longer period of time for cells in a given compartment.   

The order of mutations determines many aspects of the life of a cancer stem cell. In general, it is safe to say that the mutational selection is a filter dynamically influenced by the order of appearance of driver mutations. During the loss of stemness and acquisition of different phenotypes, a cancer cell experiences different environments, and 
in each of them, it has a different value of fitness depending on the presence of specific mutations. These particular mutations may have already occurred in the history of the cancer clone so to assure the survival. We believe that the fitness of a cancer cell is a complex function that is modified by environmental factors such as immune system, drugs and therapies. 
The introduction of the 
order of driver mutations does not completely constraint the evolution of the system; indeed, in terms of single  cells, the presence of this order does not at all affect the appearance of a mutation and is absolutely not perceived by the cells meaning that cells cannot use the right order of mutations to make any present choice or forecast any future state of the system. Therefore, single cells  will continue to acquire mutations randomly. The cells and their progenies will explore the mutation space following different trajectories not limited by a unique mutation order. Eventually, the trajectory will feel the effects of the mutation order in terms of dynamics.

The interactions between a cell and the surrounding environment is what determines the dynamics of the cell itself \cite{laplane}. As approximation, in average, a compartment is uniform and ready to interact with those cells which present the right membrane proteins. On the other hand, the cells, most of the time, are neutral to the environment interactions similar to a neutral molecule in an electric field.

We consider that driver mutations become effective only if they are preceded by other specific driver mutations. In other words, we introduce a unique sequential order of the driver mutations. If a driver
mutation is acquired in a different order from the one predefined, it remains ineffective until all the driver mutations, which should have been occurred before according to the sequential order, are activated.

\subsection{Cancer dynamics and evolution}

In order to mimic the tumour evolution, we have used an approach based on the framework of cellular automata. Each cancer cell will evolve following four types of actions depending on the internal state of the cell and the environment constraints.
The actions are asymmetric proliferation, symmetric proliferation, apoptosis and migration. These actions can be expressed in reaction form as follow:
\numparts
\begin{eqnarray}
C_{\{\dm,\Nmm,\Nndm,\co \}}&\rightarrow C_{\{\dm+\um{i},\Nmm+1,\Nndm,\co \}}+C_{\{\dm,\Nmm+1,\Nndm,\co\}},\label[equation]{reaction1}\\
\ajustspaceandequationnumber
C_{\{\dm,\Nmm,\Nndm,\co\}}&\rightarrow C_{\{\dm,\Nmm+1,\Nndm+1,\co\}}+C_{\{\dm,\Nmm+1,\Nndm,\co\}},\label[equation]{reaction1b} 
\end{eqnarray}
\endnumparts
\ajustspaceandequationnumber
\numparts
\begin{eqnarray}
C_{\{\dm,\Nmm,\Nndm,\co\}}&\rightarrow C_{\{\dm+\um{i},\Nmm+1,\Nndm,\co\}}+C_{\{\dm+\um{i},\Nmm+1,\Nndm,\co\}},\label[equation]{reaction2}\\
\ajustspaceandequationnumber
C_{\{\dm,\Nmm,\Nndm,\co\}}&\rightarrow C_{\{\dm,\Nmm+1,\Nndm+1,\co\}}+C_{\{\dm,\Nmm+1,\Nndm+1,\co\}},\label[equation]{reaction2b} 
\end{eqnarray}
\endnumparts
\ajustspaceandequationnumber
\begin{eqnarray}
C_{\{\dm,\Nmm,\Nndm,\co\}}&\rightarrow 0,\label{reaction3} 
\end{eqnarray}
\ajustspaceandequationnumber
\begin{eqnarray}
C_{\{\dm,\Nmm,\Nndm,\co\}}&\rightarrow C_{\{\dm,\Nmm,\Nndm,\co+1\}},\label{reaction4}
\end{eqnarray}

where $C_{\{\dm,\Nmm,\Nndm,\co\}}$ is the cell having  $\dm$ specific driver mutations, an amount of $\Nmm$ metabolic mutations, $\Nndm$ non--driver mutations and  being located inside the compartment $\co$. The reactions \eref{reaction1}, \eref{reaction1b}, \eref{reaction2}, \eref{reaction2b}, \eref{reaction3} and \eref{reaction4}  indicate that the cells $C$ on the left side of $\rightarrow$ undergo a process which will end with their annihilation and the creation of new cells specified by the right side of $\rightarrow$. The states of the cells are expressed between the brackets $\{\ldots\}$.

The vector $\dm$ has a dimension $\Mdm$ and belongs to the space $\{0,1\}^{\Mdm}$. Each  component $\dmc{j}$ of the vector $\dm$ takes into account the state of the single driver mutation $j\in [1, \Mdm]$, and  $\dmc{j}=1$ if the mutation is present or $\dmc{j}=0$ otherwise. 
The set of vectors $\um{i}$ have all the components equal to zero except the $i$-th component  which is equal to one. 
The number of the acquired and unordered metabolic and non--driver mutations are represented by the integer numbers $\Nmm$ and $\Nndm$, respectively.  
In the model proposed, we impose that each mutation can occur at most one time independently on the type and severity of the mutation. This persistent accumulation effect on the mutation space is more explicit for the driver mutations. In fact, the initial state of the driver mutations for any cells is given by the vector $\dm=\dmz$ having all the components equal to zero, and after a driver mutation occurs, the corresponding component become equal to one. In the same way, a further mutation relative to the same gene is excluded, meaning  the component will never switch back to zero, and that gene will never be chosen again among the possible mutating genes, hence reducing the mutation space. It is worth to remark the state of the system changes only when there is a mutation corresponding to a single gene, and its dynamic is sensible only to such mutations.  Despite the previous statements, the model can be easily extended to include a set of genes for each driver mutation. These sets do not need to be disjoint, but they can share some common genes, and when a mutation for a specific gene occurs, its state (or counting) is changed in each set where the gene is included, see \sref{operators}. Adding the above model extension will allow us to trace more precisely the diffusion in the metabolic mutation space caused by each driver mutation and their temporal order.

The reactions \eref{reaction1} and \eref{reaction1b} describe the proliferation of cells principally characterized by stem-like behaviours, hence following an asymmetric mytotic cycle. The two reactions represent the cases when a driver 
and a non--driver mutation occurs, respectively.
The reactions \eref{reaction2} and \eref{reaction2b} show the symmetric proliferation when a driver and a non--driver mutation occurs, respectively. The reactions \eref{reaction2} and \eref{reaction2b} are related to more differentiated cells which show progenitor-like characteristics.
The reaction \eref{reaction3} represents cells undergoing apoptosis. Generally this is the fate followed by cells with high instabilities or committed cells with low stem-like characteristics, which have a reduced proliferation activities even when stimulated. The last reaction, \eref{reaction4}, takes into account the migration of cells which have developed mutations allowing them to move and pass from an environment to another. The migration process depends on the occurrence of proper driver mutations which need to be developed in the proper order given by the order of the environments,  $\co \rightarrow \co+1$.

We have assumed driver and non--driver mutations are two disjoint classes of mutations, but both can be included in the wider class of the metabolic mutations. Hence, in \eref{reaction1}, \eref{reaction1b}, \eref{reaction2} and \eref{reaction2b} at each occurrence of a driver or non--driver mutation, there is an increase of one of the number of metabolic mutations. The reasons for this choice are that, on one side, the metabolic mutations mark the advancing age of the cell in terms of how many mutations steps it has acquired (see below for connections with stemness and commitment), and on the other side, the advantages and disadvantages produced by the mutations can be considered independent from the speed of the cell metabolic activity. In order to simplify  the complex representation of the underlying biological system, we do not consider reactions which result in the increasing of only metabolic mutations.  

We have six reactions when indeed there are only four distinct actions occurring in the system. The cause for these over number of reactions is related to the choice of representing the system as if there where multiple classes of mutations, and each mutation can be definitely tagged as belonging to one and only one of these classes. However, from a biological and experimental point of view, probably, not all the mutations involved can be marked with a distinct colour, but more like if they were a mix of shades. Indeed, a gene (and the molecules derived from its transcription and translation) can have a function in a compartment or at a given state and play another role somewhere else. 
For the sake of simplicity, we have grouped all the non--driver mutations in an integer variable with the purpose of counting them and neglecting the order of their occurrence. First, as we said above, non--driver mutations are not strictly necessary for the worsening and the progress of the disease, but are mutations bringing small disadvantages to the cells when compared with the large advantages of the driver  mutations. Second, the order of occurrence of the non-driver mutations in respect to all the mutations is less important than for the driver mutations, because they are weakly linked to the driver mutations and to the environment fitness constraints as well as compartment transition rates.

The rates of the reactions 
\eref{reaction1}, \eref{reaction1b}, \eref{reaction2}, \eref{reaction2b}, \eref{reaction3} and \eref{reaction4} 
are not constant, but depend on the cancer stemness. Therefore, each cell has specific probabilities of proliferating or undergoing apoptosis. If $r\in[1,4]\subset N$ is the index addressing one of the four reactions defined above, then we can define the probability function $P_r(\cs)$ that the reaction $r$ occurs as follow:
\begin{eqnarray}
	P_r:\cs\in \mathit{S}\rightarrow [0,1], &  \\
	\sum_r P_r(\cs)=1                 &  & \forall \cs,
\end{eqnarray}
where $\cs$ is a real positive value in $[0,1]$ and it represents the cancer stemness of the cell. The stemness can be generated by a random process.

The cancer stemness does not depend explicitly on the gene expression, but only on the mutation state.
Therefore, on the complete mutational space, the cells 
proceed along trajectories of cancer stemness which increases or decreases due to acquisition of new driver and new non--driver mutations, while the stemmness does not depend directly on the number of metabolic mutations. 
Eventually, if cells survive, they will be attracted toward the asymptotic value given by $\cs(\max(\Nem),\max(\Nndm))$. 
The amount of cells per compartment is limited by a carrying capacity which determines the maximum number of cells that can be stored in a given compartment. The maximum number of cells depends on the oxygen/energetic needs and it changes in function of the type of the metabolic activities. The cumulative energetic needs are computed as:
\begin{eqnarray}
E=\sum_j e(j) N(j)
\end{eqnarray}   
where $j$ is the number of driver mutations accumulated, 
$e$ is the energetic needs depending on the accumulated driver mutations and $N$ is the total number of cells in a compartment with given $j$ mutations.

\subsection{Master equation of the mutation process}

Let us consider the cell state $\stt$  with $\dm$ driver mutations, $\Nndm$ passenger mutations,  $\Nmm$ metabolic mutations and in the compartment $\co$. the index total $\Nmm$  represents also the age of the trajectory in the natural time of the occurrence of the events  $\Nmm= \dm \cdot \dm +\Nndm$.

The rates of transition involved in each action depends on the cancer cell stemness which is linked
to the internal state of the cell, $\stt$, and to the number of mutations, $\Ndm$ and $\Nndm$. Also the probabilities of acquiring a driver or a passenger mutation depend on the number of mutations already accumulated by the cells.

The conditional probability density $\rho(\dm,\Nndm,\Nmm)$ that a cell starting from the state $\sttz$ arrives after $\Nmm$ steps to the state $\stt$ is:

\begin{dmath}{\label[equation]{natural}}
\rho(\dm,\Nndm,\Nmm,\co)=
\\
\left[\frac{1}{\Mdm}\sum_{i=1}^\Mdm \ras(\Ndm-1,\Nndm,\co) \rdm(\Ndm-1,\Nndm,\co) \rho(\dm-\um{i},\Nndm,\Nmm-1,\co)\\
+\ras(\Ndm,\Nndm-1,\co)\
 \rndm(\Ndm,\Nndm-1,\co)\ \rho(\dm,\Nndm-1,\Nmm-1,\co)\right]
+2\left[\frac{1}{\Mdm}\sum_{i=1}^\Mdm \rsy(\Ndm-1,\Nndm,\co)\ \rdm(\Ndm-1,\Nndm,\co)\ \rho(\dm-\um{i},\Nndm,\Nmm-1,\co)+\rsy(\Ndm,\Nndm-1,\co)\ \rndm(\Ndm,\Nndm-1,\co)\ \rho(\dm,\Nndm-1,\Nmm-1,\co)\right]
-\rsy(\Ndm,\Nndm,\co)\ \rho(\dm,\Nndm,\Nmm,\co)-\rap(\Ndm,\Nndm,\co)\rho(\dm,\Nndm,\Nmm,\co)+[\rpa(\Ndm,\Nndm,\co-1)\rho(\dm,\Nndm,\Nmm,\co-1)-\rpa(\Ndm,\Nndm,\co)\rho(\dm,\Nndm,\Nmm,\co)]  
\end{dmath}
The first term enclosed in square brackets describes the increase of number of cells in the state $\stt$ due to asymmetric proliferation. The mutation occurring during the asymmetric proliferation can be driver or passenger; hence $r_d(\sigma)+r_n(\sigma)=1$.
The second term enclosed in square brackets takes into account the increase of cells due to symmetric proliferation, while the third term express the fact that both the daughter cells equally change their state and there is no self-renewal. 
The fourth term is the decreasing of cells due to apoptosis, and the last term is due to the change of compartment.
For the sake of simplicity, from now on, we will represent the rates without explicit dependence on the cell internal state whenever possible. 

To have a description of the dynamics comparable with the biological process, we need to switch from the natural time of the events,  regulated by the internal clock of each single unit (the cell) advancing only when an event at that scale occurs, to the physical time when the events actually occur \cite{henry}, meaning when the time is measured with a macroscopic clock regularly advancing  after a large and in average constant amount of events happen. To do so, we introduce the conditional probability density $\rho(\dm,\ndm,i+1)$ that a cell starting a time $t=t_0$ in the state $\sttz$ and exactly at time $t$ the cell changes for the $(i+1)$-th time its state to be equal to $\stt$:

\begin{dmath}
\rho(\dm,\Nndm,\Nmm,\co,t)=\int_0^t\left\{\psi(\Nmm-1,t-t')\\
\left[\left(\frac{1}{\Mdm}\sum_{i=1}^\Mdm \ras\ \rdm \ \rho(\dm-\um{i},\Nndm,\Nmm-1,\co,t')+\ras\ \rndm \rho(\dm,\Nndm-1,\Nmm-1,\co,t')\right)
+2 \left(\frac{1}{\Mdm}\sum_{i=1}^\Mdm \rsy\ \rdm\ \rho(\dm-\um{i},\Nndm,\Nmm-1,\co,t')+\rsy\ \rndm\ \rho(\dm,\Nndm-1,\Nmm-1,\co,t')\right)\right]
+\psi(\Nmm,t-t')\left[\vphantom{\sum_{i=1}^\Mdm}-\rsy\ \rho(\dm,\Nndm,\Nmm,\co,t')-\rap\ \rho(\dm,\Nndm,\Nmm,\co,t')+
\left(\vphantom{\sum_{i=1}^\Mdm}\rpa\ \rho(\dm,\Nndm,\Nmm,\co-1,t') -\rpa\ \rho(\dm,\Nndm,\Nmm,\co,t')\right) 
\right]\right\} \de t',
\end{dmath}
where $\psi(i,t)$ is the waiting time distribution that after $i$ events,  the next event occurs exactly at time $t$. The integral means we consider that the event before the last may have occurred at any possible time $t'$ between $0$ and $t$. The waiting time depends on the number $i$ of mutations occurred. The corresponding survival distribution is given by the relation:
$$\Psi(i,t)=1-\int_{0}^{t}\psi(i,t')\ \de t'.$$

If we consider an ensemble of cells, whose genome generates a trajectory in the mutation state which starts from a common initial state (the healthy state) and evolves in time.
If we let the system run and then freeze it, what we see is a population of cells which are in different states. Differently from the natural time frame, we observe cells which have accomplished unequal amounts of jumps; therefore, if $i$ is associated to the age of the cells, then at any time $i$ in the natural time frame, the system is composed of cells having the same age, while at any instant in the physical time frame, there are cells with few mutations together with cells which have accumulated a large number of mutations.

The total amount of cells with $\stt$  mutations observed at time $t$ are those cells which changed their state to  $\sigma$ (with any possible order of the occurrence of $\dm$ and $\ndm$) at an earlier time $t'$ without further events between the time $t'$ and $t$ included jumping exactly at time $t$. Hence the probability density $p(\dm,\Nndm,\Nmm,\co,t)$ of finding a cell in $\sigma$ is: 

\begin{dmath}{\label[equation]{probdens}}
p(\dm,\Nndm,\Nmm,\co,t)= \int_0^t \rho(\dm,\Nndm,\Nmm,\co,t')\ \Psi(\Nmm,t-t')\ \de t'= \delta(\dm-\dm_{0})\delta(\Nndm-\ndm_{0})\Psi(\Nmm,t)+\int^t_0  \rho^+(\dm,\Nndm,\Nmm,\co,t')\Psi(\Nmm,t-t')\ \de t,
\end{dmath}
where we have performed  a Riemann-Stieltjes integral over time with a discontinuity of the conditional probability densities $\rho$ in $t=0$ which can be expressed as: 
\begin{dmath}
{\label[equation]{discontinuity}} 
\rho(\dm,\Nndm,\Nmm,\co,t')=  \delta(\Nndm-\ndm_{0})\delta(t-0^+)+\rho^+(\dm,\Nndm,\Nmm,\co,t').
\end{dmath}

The time derivative of the previous equation give:
\begin{dmath}
\frac{\de}{\de t}\ p(\dm,\Nndm,\Nmm,\co,t)=
\rho^+(\dm,\Nndm,\Nmm,\co,t)-{\int_0^t\rho^+(\dm,\Nndm,\Nmm,\co,t')\psi(\Nmm,t-t') \de t'-\delta(\dm-\dm_{0})\psi(\Nmm,t)}
\end{dmath}

where the flux of particle exiting the $\sigma$ state is:
\begin{dmath}
{\label[equation]{outflux}}
j(\dm,\Nndm,\Nmm,\co,t)=
\int_0^t\rho(\dm,\Nndm,\Nmm,\co,t')\psi(\Nmm,t-t') \de t'
=
\int_0^t\rho^+(\dm,\Nndm,\Nmm,\co,t')\psi(\Nmm,t-t') \de t'+\delta(\dm)-\dm_{0})\psi(\Nmm,t).
\end{dmath}

Therefore, the master equation in terms of incoming and outgoing fluxes are:
\begin{dmath}
\frac{\de}{\de t}\ p(\dm,\Nndm,\Nmm,\co,t)=\rho^+(\dm,\Nndm,\Nmm,\co,t)-j(\dm,\Nndm,\Nmm,\co,t).
\end{dmath}

Using \eref{discontinuity}, we can write the incoming flux in terms of the outgoing flux defined in \eref{outflux}. The explicit result for the specific set of reaction is: 
\begin{dmath}
\rho^+(\dm,\Nndm,\Nmm,\co,t)=
\left\{\\
\left[\frac{1}{\Mdm}\sum_{i=1}^\Mdm \ras\ \rdm \ j(\dm-\um{i},\Nndm,\Nmm,\co,t)+\ras\ \rndm j(\dm,\Nndm-1,\Nmm,\co,t)\right]\\
{+2 \left[\frac{1}{\Mdm}\sum_{i=1}^\Mdm \rsy\ \rdm\ j(\dm-\um{i},\Nmm,\co,t)+\rsy\ \rndm\ j(\dm,\Nndm- 1,\Nmm,\co,t)\right]}
-\rsy\ j(\dm,\Nndm,\Nmm,\co,t)-\rap\ j(\dm,\Nndm,\Nmm,\co,t)+[\rpa\ j(\dm,\Nndm,\Nmm,\co-1,t)-\rpa\ j(\dm,\Nndm,\Nmm,\co,t)] 
\right\}.
\end{dmath}

In order to express the master equation only in terms of the probability density (see the derivation in \cite{fedotov} for more general case), we can Laplace transform both \eref{probdens} and \eref{outflux}:
$$\laplace\{p(\dm,\Nndm,\Nmm,\co,t)\}=\widetilde{p}(\dm,\Nndm,\Nmm,\co,s)= \laplace\{\rho(\dm,\Nndm,\Nmm,\co,t)\}\ \laplace\{\Psi(\Nmm,t)\},$$
$$\laplace\{j(\dm,\Nndm,\Nmm,\co,t)\}=\widetilde{j}(\dm,\Nndm,\Nmm,\co,s)= \laplace\{\rho(\dm,\Nndm,\Nmm,\co,t)\}\ \laplace\{\psi(\Nmm,t)\},$$
from which it is easy to derive the relation between the transformed probability density $\widetilde{p}$ and the transformed influx of cells $\widetilde{j}$ in terms of the transformed memory kernel $\widetilde{K}$:  
\begin{equation}
\widetilde{j}(\dm,\Nndm,\Nmm,\co,s)= \widetilde{p}(\dm,\Nndm,\Nmm,\co,s)\frac{\widetilde{\psi}(\Nmm,s)}{\widetilde{\Psi}(\Nmm,s)},
\end{equation}
and 
\begin{equation}
\widetilde{K}(\Nmm,s)=\frac{\widetilde{\psi}(\Nmm,s)}{\widetilde{\Psi}(\Nmm,s)}.
\end{equation}

The previous results allow us to rewrite the master equation in terms only of the Laplace transform of the probability density:

\begin{dmath}\label{laplaceME}
{\laplace\left\{\frac{\de}{\de t}\ p(\dm,\Nndm,\Nmm,\co,t)\right\}=s\ \widetilde{p}(\dm,\Nndm,\Nmm,\co,s)-p(\dm,\Nndm,\Nmm,\co,0)}=
\widetilde{K}(\Nmm-1,s)\\
\left\{ 
{\left[\frac{1}{\Mdm}\sum_{i=1}^\Mdm \ras\ \rdm\ \widetilde{p}(\dm-\um{i},\Nndm,\Nmm-1,\co,s)+\ras\ \rndm \ \widetilde{p}(\dm,\Nndm,\Nmm-1,\co,s)\right]}
\right.
+\left. 
2 {\left[\frac{1}{\Mdm}\sum_{i=1}^\Nmm \rsy\ \rdm\  \widetilde{p}(\dm-\um{i},\Nndm,\Nmm-1,\co,s)+\rsy\ \rndm\ \widetilde{p}(\dm,\Nndm,\Nmm-1,\co,s)\right]}
\right\}\\
+\widetilde{K}(\Nmm,s)
\left\{
\vphantom{\sum_{i=1}^\Nmm}
-\rsy\ \widetilde{p}(\dm,\Nndm,\Nmm,\co,s)-\rap\ \widetilde{p}(\dm,\Nndm,\Nmm,\co,s)+
\left[\vphantom{\sum_{i=1}^\Nmm}\rpa\ \widetilde{p}(\dm,\Nndm,\Nmm,\co-1,s)-\rpa\ \widetilde{p}(\dm,\Nndm,\Nmm,\co,s) \right]
\right\}.
\end{dmath}
The master equation can be directly anti-Laplace transformed resulting in integro-differential equations with memory kernel $K(\Nmm,t)$. Nevertheless, explicitly introducing  the functional form of the waiting time distributions may result in further simplifications.     
\subsection{Metabolic waiting time distributions}
In order to take into account the effect of acceleration due to metabolic mutations, we introduce an exponential waiting time which depends on the number of metabolic mutations $\Nmm$:
\begin{dmath}
\psi(\Nmm,t)=\alpha(r+\Nmm) e^{-\alpha r t}\ e^{-\alpha \Nmm t}, \label{waiting}
\end{dmath}
where $\frac{1}{\alpha r}$ is the cell cycle mean time of cells with no mutations, and 
$\alpha$ is the increase of the cell cycle rate caused by
each metabolic mutation.
With this choice, the Laplace transform of the memory kernel is  
$\widetilde{K}(\Nmm,s)=\alpha(r+\Nmm)$. Substituting  the previous result in the Laplace transform of \eref{laplaceME}
and anti-Laplace transforming, we obtain the final form of the master equation: 
\begin{dmath}
\frac{\de}{\de t}\ p(\dm,\Nndm,\Nmm,\co,t)= 
{\alpha (r+\Nmm-1)} \\
\left\{ 
{\left[\frac{1}{\Mdm}\sum_{i=1}^\Mdm \ras\ \rdm\ p(\dm-\um{i},\Nndm,\Nmm-1,\co,t)+\ras\ \rndm \ p(\dm,\Nndm,\Nmm-1,\co,t)\right]}
+2 {\left[\frac{1}{\Mdm}\sum_{i=1}^\Nmm \rsy\ \rdm\  p(\dm-\um{i},\Nndm,\Nmm-1,\co,t)+\rsy\ \rndm\ p(\dm,\Nndm,\Nmm-1,\co,t)\right]}\right\}\\
+{\alpha(r+\Nmm)}\left\{\vphantom{\sum_{i=1}^\Nmm}
-\rsy\ p(\dm,\Nndm,\Nmm,\co,s)-\rap\ p(\dm,\Nndm,\Nmm,\co,t)\right.+
\left.\left[\vphantom{\sum_{i=1}^\Nmm}\rpa\ p(\dm,\Nndm,\Nmm,\co-1,t)-\rpa\ p(\dm,\Nndm,\Nmm,\co,t) \right]
\right\}.\label{masterequation}
\end{dmath}

\subsection{Equations and range of parameters }

To completely describe the evolution of cancer cells' probability densities given in \sref{masterequation}, we need to define the transition rates for each type of action which depend on the mutational state of the cells in such a way to mimic the effects of the cancer stemness and the order of driver mutations. Therefore, we introduce a sufficiently generic form valid for all 4 type of actions described in \eref{reaction1}, \eref{reaction1b}, \eref{reaction2}, \eref{reaction2b}, \eref{reaction3} and \eref{reaction4} which includes an explicit dependency on the cancer stemness $\cs$ and on the amount of effective driver mutations $\Nem$:   
\begin{dmath}
	\label{transitions}
r_{\act}=\frac{1}{N}\ A_{\act}\ \chi_{\act}(\cs)\ k_{\act}(\Nem,\co).
\end{dmath}

The rates are given by the product of a constant rate amplitude $A_{\act}$ modulated by a support function $\chi_{\act}(\cs)$ depending only on the stemness of a cell multiplied by a filter function $k_{\act}(\Nem,\co)$ which is a function of the number of effective driver mutations and the compartment $\co$ divide by the normalization factor $N$. Each of the terms in the rate are action type dependent.
The amplitude $A_{\act}$ determines how much more recurrent is the specific action in a given compartment. The support function $\chi_\act\in[0,1]$ has the purpose of determining which actions are active for a specific value of the stemness $\cs$. The filter $k_{\act}\in\{0,1\}$ selects the subgroup of cell states which present specific properties to perform an action. 
For each cell state, the normalization factor is defined as:
$$
N=\sum_\act r_\act.
$$ 

Because the cancer stemness plays  a role in the survival and proliferation capability of the cells and is less important in the transition from one tissue to another, we consider
$\chi_{\mathrm{pass}}=1$. Vice versa the effective driver mutations are mainly relevant for the migration of cells from one compartment to another, hence
$k_{\mathrm{asym}}=k_{\mathrm{sym}}=k_{\mathrm{apop}}=1$.
For the remaining modulating functions, if the actions are allowed only for one contiguous compact interval of the cancer stemness $\cs$, then a valid choice for the support is:
$$
\chi_{\act}(\cs)=\Theta(\cs-w/2+c_{\act})\ \Theta(w/2-\cs-+c_{\act}),
$$
where $\Theta$ is a step function, and  $w$ is the extension of the support centred around $c_{\act}$.

The former condition can be relaxed and more phenomenological derived  functional forms, which do not present discontinuities and are derivable, can be chosen as support functions. Nevertheless, the product of Heaviside functions makes the problem simpler and allow us to describe a large variety of relevant cases.

To obtain the effect of asymmetric division for more stem-like cells and apoptotic tendency for more committed cells, then:
$$c_{\mathrm{asym}}<c_{\mathrm{sym}}<c_{\mathrm{apop}}<\overline{\cs}.$$   
These non-holonomic constraints, do not imply the support functions belong to disjoint range of the cancer stemness domain. Indeed, they can overlap so to mimic more realistic biological cases 
where for the same value of cancer stemness corresponds to multiple possible actions. On the other hand, for  disjoint support functions, the cancer stemness strictly dictates the cell behaviour.

The filter function $k_{\mathrm{pass}}(\Nem,\co)$ is a simple step function which is zero if the cell does not have all the necessary driver mutations  $S_\co$ defined in \eref{2primed}, and it is 1 if $\Nem\geq S_\co$. 
It is worth to remark that effective driver mutations can be derived from the vector of driver mutations $\dm$ by applying ladder operators similar to those used in quantum mechanics for  describing the energetic state of a harmonic oscillator which automatically accounts for the order of driver mutations (for more details see \sref{operators}).    

The cancer stemness is a positive real value function in $[0,\overline{\cs}]$ defined on the sub-space of driver and non--driver
mutations. As discussed in 
\sect{stemness},
the cancer stemness is related to the population of cancer cells due to their fitness. 
While the driver mutations (and the respective gene expressions) are responsible for the production of proteins and receptors on the surface of the cell membrane, the non--driver mutations are responsible for the cell instabilities related to the cytoskeleton. Hence a possible choice for the cancer stemness is 
$\cs(\Ndm,\Nndm)=\overline{\cs}\frac{\Ndm}{\sqrt{1+\Nndm}}$.
The surface produced by $\cs$ is a strictly monotonically  increasing function of the driver mutations $\Ndm$ and a strictly monotonically decreasing function of the non--driver mutation $\Nndm$, while it does not depend on the metabolic mutations $\Nmm$.  The variation of the cancer stemness for each cell is a random process driven by the acquisition of mutations. If we regard the $\Ndm$ and $\Nndm$ as continuous parameters is it possible to define isocurves of cancer stemness. Cells whose trajectories remain on the same isocurve while acquiring mutations will not change their (average) behavior,  while cells crossing the isocurves  will become more stem--like if going toward higher levels or more committed if going toward lower levels.

\crefalias{section}{appsec}

\renewcommand{\statevec}[1]{\boldsymbol{#1}}
\renewcommand{\operate}[1]{\left( {#1} \right)}
\renewcommand{\inner}[2]{\left\langle \statevec{#2},#1 \right\rangle}

\renewcommand{\mutM}{M}
\renewcommand{\maskM}{D}
\renewcommand{\mutV}{m}
\newcommand{\dys}{dysfunction operator}
\renewcommand{\dys}{deregulated pathway operator}

\subsection{Operators}\label{operators}

\subsubsection{Mutation operator and order of driver mutations}

If we define the vector $g=\{g_1,\ldots,g_\Ndm\}$ where $g_i$ is the $i$-th gene
and the operator $\pcheck_g$ such that given a set of genes $S$, $\pcheck_g(S)$ returns a vector 
$\statevec{\eta} =\{\eta_1,\ldots,\eta_\Ndm\}$ with components 
\begin{equation}
\eta_i=\cases{
	1 & if $\ g_i\in S$\\
	0 & otherwise},
\end{equation}
then applying $\pcheck_g$ to each element of the sequence $S^\prime$ defined in \eref{primed}, we obtain 

$$
C=\pcheck_g\circ\, S^\prime= \{\pcheck_g(S^\prime_1),\ldots,\pcheck_g(S^\prime_\Mco)\}=\{\statevec{\eta}_1,\ldots,\statevec{\eta}_\Mco\}.
$$ 
$C$ is the matrix in $[0,1]^{\Mco\times \Mdm}$ describing which mutated gene is considered necessary in a given compartment where the rows $\statevec{\eta}_\co$
represent the genes  and the columns $\grille_j$ identify a gene in the compartments. 
One or more genes may be important in various compartments and $C$ may not be a full rank matrix.

Let us define a state vector $\mutjevec \in \mathbb{R}^2$
corresponding to the gene $ g_j$ in the mutations representation so that, in the standard 
basis 
$\mathcal{B}_{\mutV_j}=\left\{\left({1 \atop 0}\right)\,\left({0 \atop 1}\right) \right\},$
we can introduce the mutation observable as a diagonal matrix operator
$$\mutjop= \matx{1& 0 \\0 & -1}$$
which acts on the respective eigenvectors as follows:
$\mutjop \operate\mutjevec= \mutjeval\mutjevec$.
The mutation operator has two eigenvalues $\mutjeval$: 1  and -1 representing the case gene $g_j$ is mutated or not mutated, respectively. The corresponding eigenvectors are equal to the basis vectors.
Generally, there are multiple genes involved in the dysfunction of a cell or in the appearance of a new oncogenic behaviour. Hence, the 
vector space of the genes' multitude can be represented as the direct product of single gene vectors: 
$$\mutevec = \bigotimes_{j=1}^\Mdm \mutjevec \in \mathcal{M}\subseteq \mathbb{R\vphantom{2^'}}^{\big({2\vphantom{2^'}}^\Mdm\big)},$$
and, similarly, the mutation operator for multiple genes is given by:
$$\mutop=\bigotimes_{j=1}^\Mdm \mutjop,$$
where, formally, the symbol $\otimes$ can be interpreted as the Kronecker product of the operators $\mutjop$. 
The mutations operator applied to an eigenvector of genes returns the eigenvector times the product of all the genes' eigenstates:   
$$\mutop \operate\mutevec = \prod_{j=1}^{\Mdm} \mutjeval \mutevec=\muteval \mutevec.$$
From the mutation operator $\mutjop$ and the elements $c_{\co j}$ of the matrix $C$ of necessary genes in a compartment, we derive the operator for a single gene:   
$$\maskjop = \id-\frac{1}{2} c_{\co j}\left( \id - \mutjop \right) =
\matx{1 & 0 \\ 0 & 1-c_{\co j}}
$$
which measures the 
deregulation
introduced by a gene in a set of important alterable genes like those belonging to a genetic pathway.
$\maskjop$ is diagonal in the mutations representation ($\mathcal{B}_{\mutV_j}$), and the two eigenvalues $\maskjeval$ 
are degenerate with value 
$1$ if the gene $g_j$ is unimportant in compartment $\co$, while they 
are non degenerate 
if the gene is
important. In the latter case, 
$\maskjeval$ is 0 when $g_j$ mutated and 1 otherwise. 
As shown before, we can use the direct product over all the genes to define the \dys{} referring to  compartment $\co$ (or to a genetic pathway) for all the involved genes: 
$$\maskop=\bigotimes_{j=1}^\Mdm\maskjop.$$

If the state vector $\mutevec$ is an eigenvector of $\mutop$, it is also an eigenvector of the \dys: 
$$\maskop \operate\mutevec=\prod_{j=1}^{\Mdm} \maskjeval \mutevec = \maskeval \mutevec,$$ and the eigenvalue $\maskeval \in \{0,1\}$ determines if all the genes in compartment $\co$ are mutated. In case we want to compare the state of two different cells or the states of a cell at two different times given by $\statevec{\mutV1}$ and $\statevec{\mutV2}$, we define the inner product 
$\inner{\statevec{\mutV1}}{\mutV2}=\delta_{\mutV1,\mutV2}$  so that:
$$ \inner{\maskop\operate{\statevec{\mutV1}}}{\mutV2} = \maskeval\ \delta_{\mutV1,\mutV2}.$$
It is straightforward noticing that the \dys{} allows us to properly address the problem of inter compartmental unordered cumulated mutations and the intra-compartmental order of mutation through the index $\co$ just by imposing the orthogonality between elements belonging to different compartments.

In fact, if $\co 1$ and $\co 2$ are two generic compartments, and $\mutevec_{\co 1}$ and $\mutevec_{\co 2}$ are the mutation vectors referring to the multiple genes in the respective compartments, then: 
$$
\operator{\maskM}_{\co 1} \operate{\mutevec_{\co 2}} =
\cases{ 
	\eval{\maskM_{\co 1}}\mutevec_{\co 1} & $\co 1=\co 2$\\
	0 & otherwise
}.$$

To easily deal with  the order of mutation among compartments in terms of operators, we need to extend the notation by defining 
the level state vector 
$\statevec{\ell} \in\mathcal{L}$ where
the vector space $\mathcal{L}$ is given by the direct sum of $\Mco$ identical $\mathcal{M}$ space vector, which for convenience we rename as $\mathcal{M}_k$, such that: 
$$\mathcal{L}=\bigoplus_{\co=1}^\Mco\mathcal{M}=\bigoplus_{\co=1}^\Mco\mathcal{M}_k\subseteq \mathbb{R\vphantom{2^'}}^{\big({2\vphantom{2^'}}^\Mdm\big)\cdot \Mco}.$$ 
An opportune basis for $\mathcal{L}$ is the standard basis where the first $2^\Mdm$ basis vectors are a complete basis for $\mathcal{M}_1$, and 
their linear combinations refer to 
the mutation states of all the genes while a cell is in compartment $\co=1$.  The basis vectors having the only component different from zero at positions going from $2^\Mdm+1$ to $2^{\Mdm+1}$ are a complete basis for $\mathcal{M}_2$, 
and their linear combinations  refer to the mutation state of all the genes while the cell is in compartment $\co=2$. The partition of the basis vectors of $\mathcal{L}$ can be similarly done for each compartment.

Let us call $\statevec{\ell}_\co$ the gene state eigenvectors of a cell in compartment $\co$ such that the components from $2^{\Mdm+\co} +1$ to $2^{\Mdm+\co+1}$ are identical to the respective  eigenvectors $\mutevec_k\in \mathcal{M}_k$ in the $\co$-th compartment.
In this representation, we can redefine and unify the set of \dys s in a simpler form:
$$
\operator{D}=\matx{
	\operator{\maskM}_1 & 0 &\cdots&0\\
	0& \operator{\maskM}_2  &\cdots&0\\
	\vdots&\vdots         &\ddots&0\\
	0 & 0 &\cdots&\operator{\maskM}_\Mco &\\
},
$$
and 
$\operator{D}\operate{\statevec{\ell}_{\co}}=\maskeval \statevec{\ell}_{\co}$. 

It is important to stress  all the state vectors defined above are all factorizable because genes and  compartments are all considered independent.

\subsubsection{Ladder operators and effective driver mutations}

The deregulated pathways operator $\operator{D}$ is diagonal in the representation of the orthonormal level basis $\mathcal{B}_{\statevec{\ell}_\co}=\{\statevec{\ell}_1,\statevec{\ell}_2,\ldots,\statevec{\ell}_\Mco\}$, and it only has two degenerate eigenvalues $\eval{\maskM}$ equal to 0 and 1; the null eigenvalue means not all the required genes involved in the regulation/compensation of a specific pathway are mutated so as to not result in further oncogenic activity, and the latter eigenvalue means the pro-oncogenic behavior is present and the cell can use it when needed.  
Let us introduce the ladder operators for the level of the path $\operator{L}^+$ and $\operator{L}^-$. The two operators act on the basis by increasing or decreasing the level respectively. Therefore, the creation operator gives $\operator{L}^+ \operate{\statevec{\ell}_\co}=\statevec{\ell}_{\co+1}$ for any integer $1\leq \co<\Mco$, and it returns 0 when applied on $\statevec{\ell}_{\Mco}$. 
In the opposite way, the action of the annihilation operator is   $\operator{L}^- \operate{\statevec{\ell}_\co}=\statevec{\ell}_{\co-1}$ for any integer $1< \co\leq\Mco$, and it gives 0 when applied on $\statevec{\ell}_1$. The ladder operators are the transpose of one another: 
$\operator{L}^+ =
\left(\operator{L}^-\right)
^{\mathrm{T}}$. 
For extended discussions on ladder operators and their applications see \cite{walczak,sakurai,zeidler},

Coupling the degenerate operator and the ladder operators such that
$\operator{B}^+=\operator{L}^+\operator{D}$ and $\operator{B}^-=\operator{D}\operator{L}^-$ 
is useful to determine which one of a succession of ordered events in the cell path represents a barrier. 
The barrier operators $\operator{B}^+$ and $\operator{B}^-$ applied to the elements of the basis vectors $\mathcal{B}_{\statevec{\ell}_\co}$ give
$$\operator{B}^+ \operate{\statevec{\ell}_\co}=\maskeval \operator{L}^+ \operate{\statevec{\ell}_\co}=\maskeval \statevec{\ell}_{\co+ 1}$$
and
$$\operator{B}^- \operate{\statevec{\ell}_\co}=\operator{D} \operate{\statevec{\ell}_{\co-1}}=\eval{\maskM_{\co-1}} \statevec{\ell}_{\co- 1}$$
if $ 1\leq \co+1 \leq \Mco$ and zero otherwise. 

In order to determine the value of the transition rate of each cell to jump from one compartment to the next, it is necessary to determine how many driver mutations are effective for the specific cell. In a system with a finite number of levels, meaning a finite number of compartments, this computation can be achieved  by using the effective ladder operators  
$\operator{E}^\pm=\frac{1}{1-\operator{B}^\pm\vphantom{\big|}}$.

The result of the effective ladder operators on the basis vector $\statevec{\ell}_\co$ is:

$$\operator{E}^\pm \operate{\statevec{\ell}_\co}=\sum_{
	{j=0 \atop 1\leq \co\pm j\leq\Mco}
} \left(\prod_{i=0}^{j}{\lambda_D}_{\co\pm i}\right) \statevec{\ell}_{\co \pm j}$$

where the factors $\left(\prod_{i=0}^{j}{\lambda_D}_{\co\pm i}\right)$ are different from zero only when all the ordered sequence of driver mutations %
previous to (in case of $\operator{E}^-$) and afterwards (in case of $\operator{E}^+$) the unknown value $\Nem$ in a given compartment $\co$  have already been acquired. 
The number of effective driver mutations $\Nem$ for a cell in compartment $\co$ is finally retrieved by the square of the norm of $\operator{E}^- \operate{\statevec{\ell}_\co}$ as follows:  

\begin{equation}
\inner{\operator{E}^+\operator{E}^- \operate{\statevec{\ell}_\co}}{\statevec{\ell}_\co}=\sum_{
	{j=0 \atop 1\leq \co\pm j\leq\Mco}
} \left(\prod_{i=0}^{j}{\lambda_D}_{\co\pm i}\right)=\Nem.\label{ordernormOP}
\end{equation}

This result can be applied to find the value of $\rpa$ of each cell in each compartment and at each simulation step by knowing the cell mutated genes.

\section{Results}

\subsection{Simulations}
The simulations are done in the framework of CA where each unit represents a cancer cell having a position $\stt$ in the cell state space composed by the mutation space and the tissue space. A simulation consists in a set of cells beginning with specific initial conditions and evolving together with the advancing of time.    
At each time, each cell continues to occupy the same position until an event occurs after which the corresponding cell moves to a new position in the state space. The evolution of cells, the creation of new others and their annihilation are given by stochastic events following the rules in \eref{reaction1}, \eref{reaction1b}, \eref{reaction2}, \eref{reaction2b}, \eref{reaction3} and \eref{reaction4}.  
There are two types of constraints for the cells. One is due to the limited extension of the cell state space given by the ranges $\{0,1\}^\Ndm$, $\{0,\dots,\Mndm\}$, $\{0,\dots,\Mmm\}$ and $\{1,\dots,\Mco\}$ which are already included in the master equation. The second constraint is a volume exclusion limit which impose a maximum capacity of cells having the same number of driver mutations. 
The update of the system during the simulations is based on the asynchronous self-clocked time scheme where each cell has an independent timer. When a cell undergoes a specific action all the resulting cells from the product of the reaction set their timers to a random period given by the waiting time distribution \eref{waiting} which depends on the cell state. 
Hence, the waiting times for the new events are chosen randomly after an event occurs and are based on the actual new cell state. This interval of time remains constant independently from other events occurring meanwhile. 
The type of action performed at the end of a time period is randomly chosen with probabilities given in \eref{transitions} which depend only on the internal state of the cell.
Because the state of a cell does not change between events, we are in a case of non competition between the reactions; therefore the type of action performed by a cell at the end of the time period can be tossed and decided at the beginning of the time period. At the end of the time period, the cell checks all the constraints and if the limits are not reached, the reaction is performed,otherwise it is put back.
Simulations with the same initial conditions and parameters are repeated multiple times. At each time, the number of cells in the same cell state are averaged over the ensemble of simulations. Eventually, the mean trajectories and the deviations from the mean are computed. In some cases the trajectories of the cells are projected into a sub--space of the entire cell state and then averaged so to reduce the number of degrees of freedom and highlight the main characteristics of the systems corresponding to the observables which are experimentally detected.
The program for the simulations has been written using the commercial software Wolfram Mathematica.

\subsection{Data analysis supporting metabolism-related mutations}

The Cosmic (http://cancer.sanger.ac.uk/) is the world's largest and most comprehensive database resource for exploring the impact of somatic mutations. Other valuable databases include The Gene Expression Omnibus (GEO, www.ncpi.nlm.nih.gov/gds) among others. 
We have developed a R program that performs statistical analysis on the Cosmic database, identifies and evaluates the functional impact of each mutation by using a combination of pathways and gene ontology approaches. The functional annotation and enrichment analysis allows to classify large lists of genes into metabolism related groups.  We have then extracted useful statistical estimators that allows to quantify the role of the metabolism-related mutations in the context of driver and non--driver mutations and the heterogeneity of proliferation rates among the cells. We have analysed the mutations with different threshold of FATHMM-MKL value which measures the impact of a mutation .
The FATHMM-MKL algorithm predicts the functional, molecular and phenotypic consequences of protein missense variants using hidden Markov models \cite{shihab}.

\begin{table}[htbp]
	\fontsize{8}{10}\selectfont
	\centering
	\caption{Go Ontologies ( from the top biological processes, molecular functions, cellular components ) of the mutations with the highest FATHMM score $>0.9$; the last section shows the relative enrichment of the mutations with the highest FATHMM with respect to a background of low FATHMM score below 0.7}
	\begin{tabular}{|m{12.5ex}|m{36ex}|m{11ex}|m{11ex}|m{12.2em}|}
		\hline
		\textbf{GO term} & \textbf{Description} & \textbf{P-value} & \textbf{FDR \qquad q-value} & \textbf{Enrichment (N,B, n, b)} \\ 		\hline
		\multicolumn{5}{|c|}{}\\ [-1.8ex]
		\cline{1-1}
		process & \multicolumn{4}{c|}{}\\ 		\hline
		GO:0090662  & ATP hydrolysis coupled transmembrane transport  & 7.30E-68 & 1.07E-63 & 14.24 (16103,65,1079,62) \\ \hline
		GO:0099131  & ATP hydrolysis coupled ion transmembrane transport  & 5.08E-62 & 3.72E-58 & 14.18 (16103,60,1079,57) \\ \hline
		GO:0044710  & single-organism metabolic process  & 1.84E-56 & 8.97E-53 & 2.66 (16103,3086,507,258) \\ \hline
		GO:0099132  & ATP hydrolysis coupled cation transmembrane transport  & 5.61E-53 & 2.06E-49 & 14.61 (16103,48,1079,47) \\ \hline
		GO:0044281  & small molecule metabolic process  & 2.11E-47 & 6.20E-44 & 3.36 (16103,1526,549,175) \\ \hline
		GO:0019752  & carboxylic acid metabolic process  & 2.44E-42 & 5.95E-39 & 4.62 (16103,750,530,114) \\ \hline
		GO:0006631  & fatty acid metabolic process  & 3.57E-41 & 7.47E-38 & 15.96 (16103,262,181,47) \\ \hline
		GO:0006637  & acyl-CoA metabolic process  & 7.30E-40 & 1.34E-36 & 34.72 (16103,82,181,32) \\ \hline
		GO:0035383  & thioester metabolic process  & 7.30E-40 & 1.19E-36 & 34.72 (16103,82,181,32) \\ \hline
		GO:0043436  & oxoacid metabolic process  & 8.82E-39 & 1.29E-35 & 4.16 (16103,855,530,117) \\ \hline
		GO:0006082  & organic acid metabolic process  & 1.20E-38 & 1.59E-35 & 4.11 (16103,872,530,118) \\ \hline
		GO:0032787  & monocarboxylic acid metabolic process  & 2.28E-38 & 2.79E-35 & 5.85 (16103,437,529,84) \\ \hline
		GO:0044255  & cellular lipid metabolic process  & 7.30E-37 & 8.23E-34 & 4.11 (16103,873,507,113)\\ 
		\hline
		\multicolumn{5}{|c|}{}\\[-1.8ex]
		\cline{1-1}
		molecular & \multicolumn{4}{c|}{} \\
		\hline
		GO:0043492  & ATPase activity, coupled to movement of substances  & 1.30E-114 & 5.69E-111 & 14.04 (16121,111,1086,105) \\ \hline
		GO:0016820  & hydrolase activity, acting on acid anhydrides, catalyzing transmembrane movement of substances  & 6.49E-102 & 1.42E-98 & 14.18 (16121,98,1079,93) \\ \hline
		GO:0042626  & ATPase activity, coupled to transmembrane movement of substances  & 1.51E-99 & 2.20E-96 & 14.16 (16121,96,1079,91) \\ \hline
		GO:0015399  & primary active transmembrane transporter activity  & 3.07E-97 & 3.37E-94 & 13.87 (16121,98,1079,91) \\ 
		\hline
		\multicolumn{5}{|c|}{}\\[-1.8ex]
		\cline{1-1}
		component & \multicolumn{4}{c|}{} \\
		\hline
		GO:0044444  & cytoplasmic part  & 1.16E-32 & 2.06E-29 & 1.33 (16121,7980,1311,866) \\ \hline
		GO:0005581  & collagen trimer  & 1.69E-26 & 1.50E-23 & 3.90 (16121,84,3100,63) \\ \hline
		GO:0016469  & proton-transporting two-sector ATPase complex  & 5.20E-22 & 3.09E-19 & 13.20 (16121,25,1075,22) \\ \hline
		GO:0005891  & voltage-gated calcium channel complex  & 4.81E-20 & 2.14E-17 & 6.14 (16121,37,2201,31) \\ \hline
		GO:0034704  & calcium channel complex  & 3.39E-18 & 1.21E-15 & 4.93 (16121,54,2241,37) \\ \hline
		GO:0005759  & mitochondrial matrix  & 6.33E-18 & 1.88E-15 & 5.21 (16121,303,470,46) \\ \hline
		GO:0098533  & ATPase dependent transmembrane transport complex  & 1.57E-17 & 4.00E-15 & 13.82 (16121,19,1044,17)\\ \hline
		GO:1904949  & ATPase complex  & 1.04E-16 & 2.31E-14 & 13.13 (16121,20,1044,17) \\ \hline
		GO:0043190  & ATP-binding cassette (ABC) transporter complex  & 1.09E-16 & 2.15E-14 & 240.61 (16121,7,67,7) \\
		\hline
		\multicolumn{5}{|c|}{}\\[-1.8ex]
		\cline{1-1}
		drivers vs non--drivers  & \multicolumn{4}{c|}{} \\
		\hline
		GO:0044238  & primary metabolic process  & 2.63E-12 & 5.94E-09 & 1.03 (10465,5247,9049,4658) \\ \hline
		GO:0044237  & cellular metabolic process  & 2.85E-12 & 5.53E-09 & 1.03 (10465,5305,9049,4708) \\ \hline
		GO:0071704  & organic substance metabolic process  & 4.81E-12 & 8.16E-09 & 1.03 (10465,5488,9049,4865) \\
		\hline
	\end{tabular}	
\label{tab:addlabel}
\end{table}
This analysis shows the importance of the metabolic mutations in the context of the mutation impact calculated with the FATHMM score.

\subsection{Discussion}
To highlight the role of effective driver mutations and the role of the order in which the driver mutations occur, we look at the differences with the unordered driver mutations dynamics in which cancer cells evolve and migrate to different tissue depending only on the number of driver mutations $\Ndm$ and not on the specificity of the corresponding genes.
We simulate the two distinct cases and compare the recovering capabilities of cancer cell populations after a drug targeting  sub--populations of cells within a specific range of the state $\stt$ 
has been administered in one compartment.
The effect of the drug is simulated by killing all the cells which match a given signature at the time of administration $t_{\drug}$. Therefore, let us define a drug target signature $\sigma_{\drug}=\{{\dm}_{\drug},{\Nndm}_{\ \drug},{\Nmm}_{\ \drug},\co_{\drug}\}$, the amount of driver mutations ${\Ndm}_{\ \drug}$ and 
the vector of driver mutations $\dm_{\drug}$ belonging to the same space of the cell driver mutations $\{0,1\}^\Mdm$, so that all the cells  immediately sent to apoptosis are those for which hold true both the relations:

\numparts
\begin{eqnarray}
\sigma_{\drug}\ \circ\ \sigma=\sigma_{\drug}\ \circ\ \sigma_{\drug},\label{drugtest1}\\
{\Ndm}_{\ \drug}\ ||\dm||_1={{\Ndm}_{\ \drug}}^2 \label{drugtest2}
\end{eqnarray}
\endnumparts

where $\circ$ is the Hadamard product and $||\cdot||_1$ is the taxi cab norm of a vector.
Then, we  observe the recovery of the populations 
after the perturbation induced by the drug affecting the target cells is applied at the same time in the ordered and unordered 
simulations.
In the simulations, we have considered the biological significant case of breast cancer cells metastasising in the bone. The process encompasses three different tissues. The mammary ducts are the primary site where cells develop malignant mutations and become tumourigenic. The second tissue is the circulatory system. Cancer cells which undergo  epithelial to mesenchymal transition will initiate to migrate and, chemoattracted to regions with higher concentration of oxygen and nutrients, eventually they will intravasate through the proximal blood vessels. There, the cancer cells, also named Circulating Tumour Cells (CTC), will travel in the circulatory system. The CTCs which are recognized as self  by the immune system will survive longer in the circulatory system, but only the cells capable of extravasating while reach the secondary sites. The third tissue is the bone. Breast cancer cells' preferred metastasisation sites are regions enriched in \tgfb, and bone tissue has one of the highest concentration of \tgfb. Bone tissue represents the statistically preferred secondary site for breast cancer cells. Nonetheless, not all cancer cells reaching the secondary site will be enough tumourigenic to generate metastases, but only those which undergo mesenchymal to epithelial transition show sufficient cancer stem-like behaviours with the potentiality to form a secondary colony.
Cell survival and capability of migrating is due to specific membrane proteins involved in breast cancer metastasising in the bone. A small subset of four genes corresponding to membrane proteins which positively correlate with the evolution of the disease and the survival of the patients has been experimentally found in \cite{asco2} and are EPCAM, CD47, CD44 and MET. Mutations of these genes correspond to driver mutations.
The EPCAM is related to the cells' connectivity of the mammary duct lumen and plays a role in the epithelial to mesenchymal transition \cite{Badve}. The CD47 is responsible for the production of membrane proteins increasing the survival of cancer cells by helping them to evade macrophage phagocytosis. 
CD44 variant isoforms are highly expressed in carcinomas of epithelial origin and relate to tumour progression and metastatic potential of some cancers. The protein CD44 bound with hyaluronic acid and sugar rich coating molecules is found on the surface of endothelial cells and tumour cells. Different amounts of CD44 changes the arrangement of the coating sugars causing the exposition of the embedded adhesion mediated selectin and integrins favouring cell extravasation \cite{Mitchell}. The MET gene regulates mesenchymal to epithelial transition, and its overexpression positively correlates with metastasis formation. 
The four driver mutations present a sequential order given by their respective functionality in the three tissues, hence the order of the genes and their respective mutations is $\{\textrm{EPCAM, CD47, CD44, MET}\}$. Using the definition in \eref{2primed}, we derive the following table:
\\
\begin{equation}
\begin{tabular}{|c|c|c|} 
\hline \rule[0ex]{0pt}{3.ex}  
 $S^{\prime\prime}_1$  & $S^{\prime\prime}_2$ &  $S^{\prime\prime}_3$\\ 
\hline \rule[0ex]{0pt}{2.5ex}  
breast  & circulatory sys. & bone  \\ 
\hline \rule[0ex]{0pt}{2.5ex}  
EPCAM  & CD47, CD44 & MET  \\ 
\hline \rule[0ex]{0pt}{2.5ex}  
0-1  & 1-3 & 3-4  \\  
\hline 
\end{tabular} 
\label[table]{tabexample}
\vspace{1em}
\end{equation}
\vspace{1em}
which puts in relation each compartment with a subset of mutations necessary to reach or seed in the next compartment, hence, defining the right order of driver mutations.
In the circulatory system, CTCs need two driver mutations to survive and reach the bone tissue. The order of mutations between CD47 and CD44 is not important and both are considered effective drivers. The last row of \tref{tabexample} shows the minimum and maximum number of effective driver mutations a cell can have in each compartment.

We have simulated three distinct cases of drug targets. The first two cases are less realistic, but are proposed to show the effect of drugs acting 
on unordered driver mutation systems, hence targeting cells with non specificity of the mutated gene.
While, the last case is a realistic example of a drug affecting cells in a specific compartment with specific driver mutations.
The drug target signature for the cases studied are:
\\

\begin{equation}
\begin{tabular}{|c|C{14ex}|C{14ex}|C{15ex}|} 
\hline \rule[1ex]{0pt}{3.ex}  
& case 1:\newline one driver  & case 2:\newline two drivers & case 3:\newline specific drivers\\ 
\hline \rule[0ex]{0pt}{2.5ex}  \rule[-1.1ex]{0pt}{2.5ex}
$\dm_{\drug}$  & $\{0,0,0,0\}$ & $\{0,0,0,0\}$ & $\{0,1,0,0\}$\\ 
\hline \rule[0ex]{0pt}{2.5ex}  \rule[-1.1ex]{0pt}{2.5ex}
${\Nndm}_{\ \drug}$  & 0 & 0 & 0  \\ 
\hline \rule[0ex]{0pt}{2.5ex}  \rule[-1.1ex]{0pt}{2.5ex}
${\Nmm}_{\ \drug}$  & 0 & 0 & 0  \\  
\hline \rule[0ex]{0pt}{2.5ex} \rule[-1.1ex]{0pt}{2.5ex}
$\co_{\drug}$ & 2 & 2 & 2\\
\hline \rule[0ex]{0pt}{2.5ex}\rule[-1.1ex]{0pt}{2.5ex} 
${\Ndm}_{\ \drug}$ & 1 & 2 & 0\\
\hline
\end{tabular} 
\label[table]{tabcases}
\vspace{1em}
\end{equation}
\vspace{1em}
All the components of the drug signature equal to zero are disregarded. Consequently, cells can have any value corresponding to each of the null components and still be a drug target.

\begin{figure}[ht]
	\centering
	\includegraphics[width=0.98\linewidth]{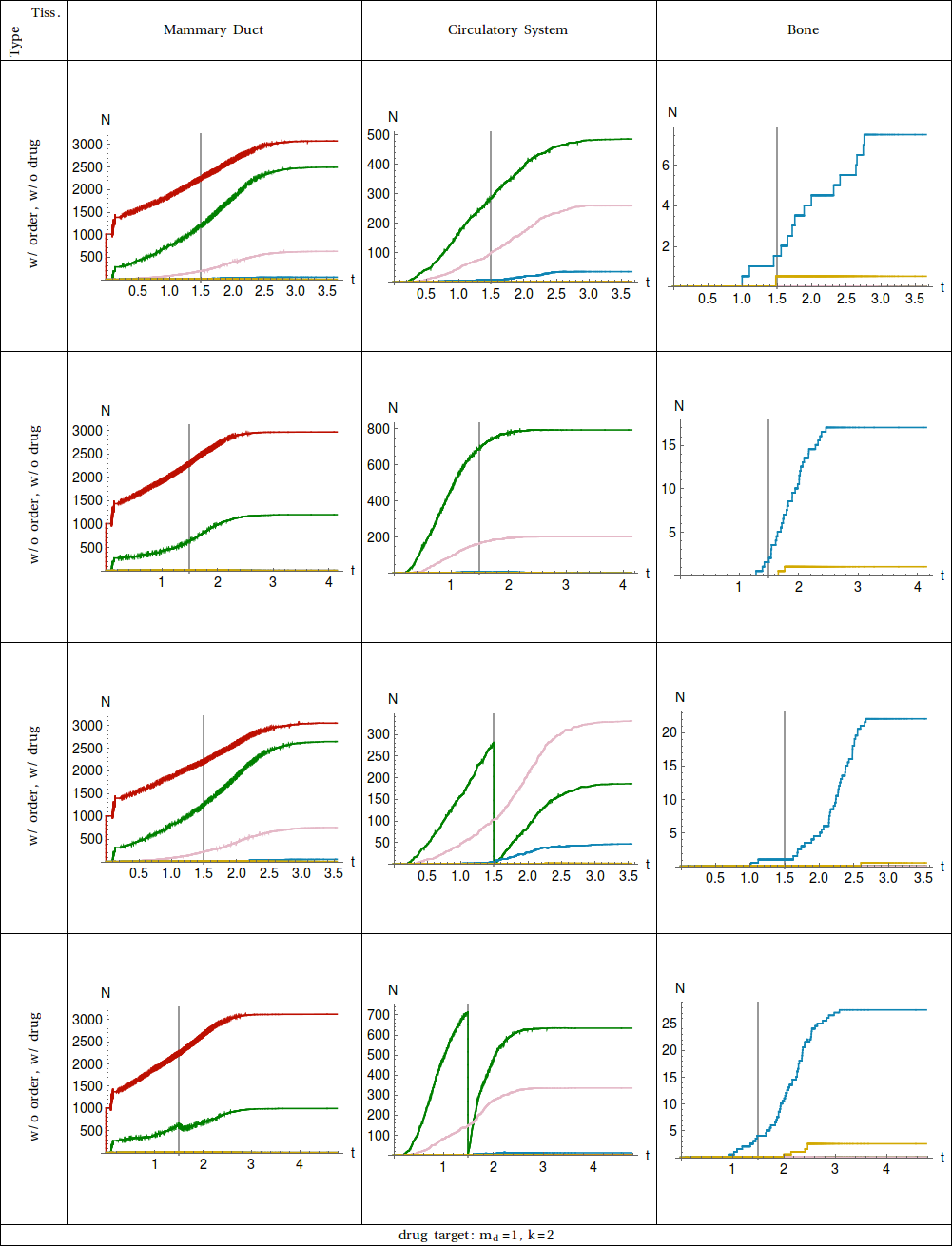}
	\caption{Plot of the number of cell populations in function of the time. The 3 tissues are shown in the columns 
	ordered from left to right following the cell traversing. All the combinations of ordered (w/ order) and unordered (w/o order) mutation dynamics together with drug (w/ drug) and without drug (w/o drug) administration are considered and shown in rows. Each curve represents the sub--population of cells with a specific number of driver mutations: $\Ndm=0$ is red, $\Ndm=1$ is green, $\Ndm=2$ is pink, $\Ndm=3$ is blue, and $\Ndm=4$ is yellow. The event of the drug administration is at time $t_{\drug}$ which is marked with a vertical line for easier comparison. The target of the drug, \eref{drugtest1} and \eref{drugtest2}, and the compartment affected is shown at the bottom.}
	\label{fig:plotmd1}
\end{figure}

In case 1 of \tref{tabcases}, all the circulating tumour cells which have acquired a single driver mutation are killed by the drug at time $t_{\drug}$ as shown in \fref{fig:plotmd1}. On the one hand, this means in the simulations with no ordered driver mutations, the cells targeted in the circulatory system could have acquired any of the four driver mutations listed in \tref{tabexample}. Considering that when the order of mutations does not affect the dynamics, the cells which have 
one of any of the driver mutations can already intravasate, and, by entering in the circulatory system, they allow the restoration of the number of CTCs with one driver 
constituting 
a reservoir for cells which later acquire a second and a third driver mutation by asymmetrical division. Furthermore, the drug does not target sub--populations of cells having two or more driver mutations. Indeed, the dynamics of the sub--populations of cells having more than one driver when the drug is administered are close to the simulations with no drugs meaning they are mainly
reservoirs of their own sub--populations due to symmetrical proliferation.

On the other hand, in simulations with ordered driver mutations, a drug targeting any cell with one driver (case 1) is more effective when compared with the simulations with no order of driver mutations only in terms of total number of CTCs. The reason for less CTCs is that the minimum requirement to intravasate is having the mutation of the EPCAM gene which is more restrictive when compared to the unordered driver dynamics. 
Indeed, the sub--population of cancer cells with two driver mutations is close to the corresponding sub--population in the simulations with no order. The sub--population of cells exiting the mummary duct and entering the circulatory system is represented in both ordered and unordered driver simulations by the sub--population with one driver mutation referring to the EPCAM gene in the former and  to a generic driver in the latter. 
Case 1 highlights a set of situations where perturbing sub--populations of cells in a compartment with few driver mutations compared to the total number of mutations necessary to change compartment, hence which are upstream in respect to the local progenies, 
does not affect sub--populations with more driver mutations, and therefore, the role of asymmetric proliferation is less relevant due to  the loss of cancer stemness.  

\begin{figure}[ht]
	\centering
	\includegraphics[width=0.98\linewidth]{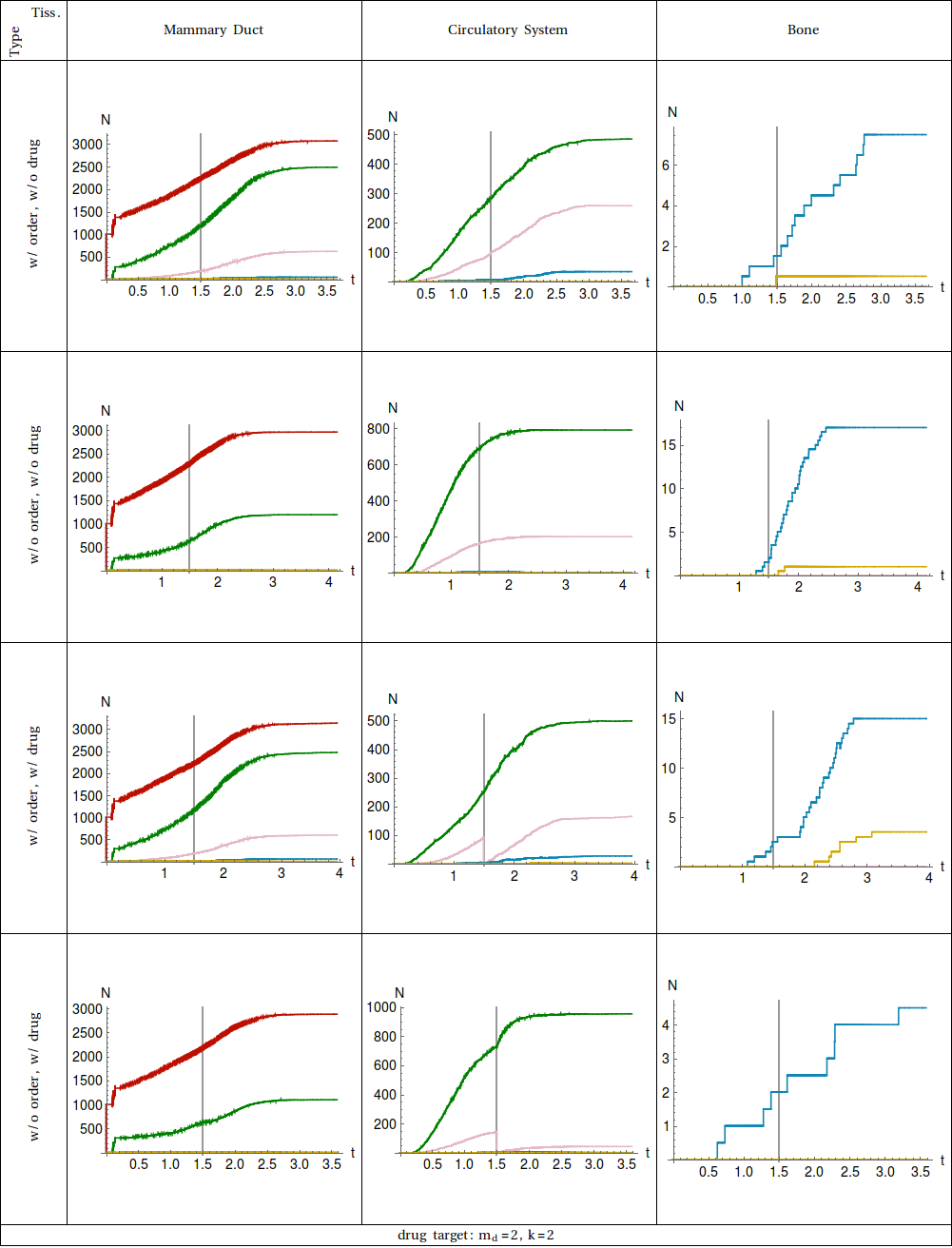}
	\caption{Plot of the number of cell populations in function of the time. The 3 tissues are shown in the columns 
	ordered from left to right following the cell traversing. All the combinations of ordered (w/ order) and unordered (w/o order) mutation dynamics together with drug (w/ drug) and without drug (w/o drug) administration are considered and shown in rows. Each curve represents the sub--population of cells with a specific number of driver mutations: $\Ndm=0$ is red, $\Ndm=1$ is green, $\Ndm=2$ is pink, $\Ndm=3$ is blue, and $\Ndm=4$ is yellow. The event of the drug administration is at time $t_{\drug}$ which is marked with a vertical line for easier comparison. The target of the drug, \eref{drugtest1} and \eref{drugtest2}, and the compartment affected is shown at the bottom.}
	\label{fig:plotmd2}
\end{figure}

In case 2 of \tref{tabcases}, the drug targets cancer cells in the second compartment with two driver mutations. Cell sub--populations are plotted in \fref{fig:plotmd2}. As in the previous case, the results present different dynamics and drug conditions in all three compartments. The effect of the treatment is equal in both simulations with order and without order of driver mutation respectively. Indeed, the sub--population of circulating tumour cells with two driver mutations are going to zero at time $t_{\drug}$. The differences between the two simulations is given by the dynamics of recovering of the targeted sub--populations and by the competition with the non targeted ones. 

Similarly to case 1, when no treatment is administered, 
the number of CTCs with one driver mutation is larger in the simulations with unordered dynamics due to a less restrictive filter on the transition rates of cancer cells to pass from the mammary duct into the circulatory system than in the simulations with ordered dynamics. On the contrary, sub--populations with two driver mutations is larger in the ordered dynamics than in the unordered one because there are larger reservoirs of cancer cells  in the mammary duct 
which have a null transition rate as a consequence of the wrong order of driver mutations. These reservoirs are composed of heterogeneous populations of cancer cells which go from committed phenotypes to more stem-like phenotypes. 
When the necessary mutations are acquired (in this specific case the required mutation is on the EPCAM gene) by any of the cells in the reservoirs, their transition rate becomes different from zero, and they contribute to the sub--populations with higher numbers of driver mutations shown in the last row of \tref{tabexample}. Cells with more stemness will have more chances to survive and divide asymmetrically, while committed cells will have less possibility to contribute to the evolution of the disease. Similarly, many CTCs are obliged to remain in circulatory system until they have all the required mutations necessary for the extravasation. Hence, in the simulations with order of driver mutations, there are CTCs with more than two driver mutations, while in the simulations without order, there is only a sub--population of cells with two driver mutations which are already ready to extravasate.

The evolution of sub--populations of cancer cells reaching the bone compartment are similar in both ordered and unordered types of simulations when the system is not perturbed at all. Instead, when a perturbation is applied at time $t_{\drug}$, the simulations with unordered dynamics show no sub--population of cells capable of seeding and metastasise in the bone, while, in the simulations with order of driver mutation dynamics, the number of aggressive cancer cells capable of reaching the bone tissue is even larger than in the absence of perturbation. This major difference between the two dynamics stems from the fact that in simulations with ordered driver mutations there are CTCs having high value of cancer-stemness which are missing only few (in this case just the CD47 and CD44 genes) driver mutations, but are constrained to remain in the circulatory system. The lack of extravasating capabilities make these sub--populations of cancer cells apparently less aggressive, when instead they are a few mutational steps away from the aggressive phenotype. Furthermore, part of these cells acquired a large number of metabolic mutations which contribute to the speed of the cell cycle and their effective capability of forming secondary colonies.

\begin{figure}[ht]
	\centering
	\includegraphics[width=0.98\linewidth]{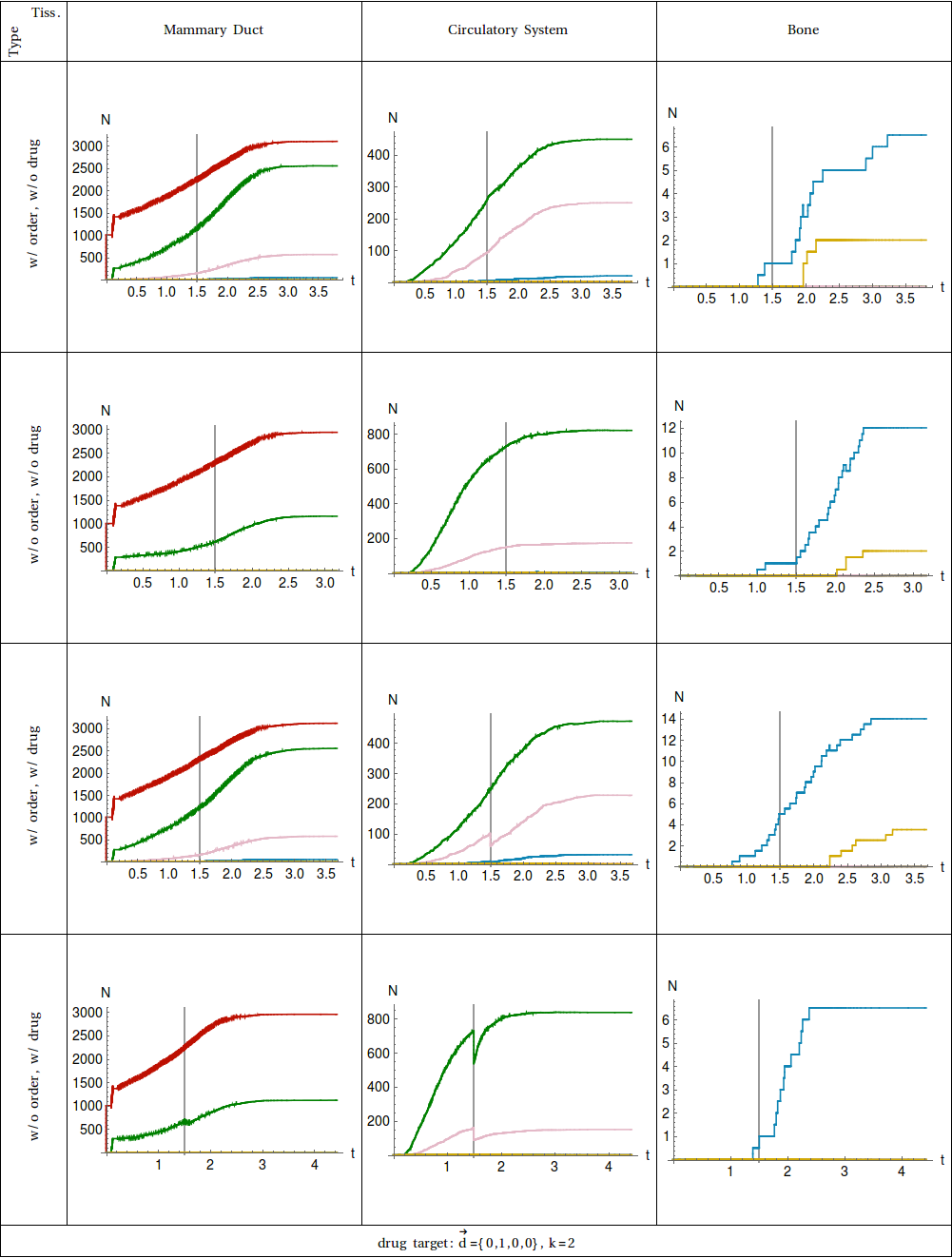}
	\caption{Plot of the number of cell populations in function of the time. The 3 tissues are shown in the columns 
	ordered from left to right following the cell traversing. All the combinations of ordered (w/ order) and unordered (w/o order) mutation dynamics together with drug (w/ drug) and without drug (w/o drug) administration are considered and shown in rows. Each curve represents the sub--population of cells with a specific number of driver mutations: $\Ndm=0$ is red, $\Ndm=1$ is green, $\Ndm=2$ is pink, $\Ndm=3$ is blue, and $\Ndm=4$ is yellow. The event of the drug administration is at time $t_{\drug}$ which is marked with a vertical line for easier comparison. The target of the drug, \eref{drugtest1} and \eref{drugtest2}, and the compartment affected is shown at the bottom.}
	\label{fig:plotd0100}
\end{figure}

Case 3 of \tref{tabcases} differentiates from the former cases for the specificity of the drug that targets only circulating tumour cells with the CD47 gene mutated, see \fref{fig:plotd0100}. In the dynamics with ordered driver mutations, the sub--populations affected are those with two and three driver mutations, but the treatment leaves untouched cells with only one driver mutation referring to the EPCAM gene. In this case, we can see the corresponding curves do not go to zero as in the former cases because the respective sub--populations contain cells without the CD47 mutation.	In the unordered dynamics, the drug also hit the sub--population of cells with one driver mutation, and  it is more effective compared with the ordered dynamics. Precisely in terms of combinatorics, the drug in the unordered dynamics targets $1/4$ of all the combinations of cells with one driver mutations, $1/2$ of the combinations of cells with two driver mutations and $3/4$ of the combinations of cells with three driver mutations. In the dynamics with ordered driver mutations, the drug targets none of the cells with one driver mutation, $1/3$ of the combinations of cells with two drivers and $2/3$ of the combinations of cells with three driver mutations.

The order of driver mutations introduces a strong constraint in the migration of cancer cells from one compartment to another compared to a cancer evolution driven by unordered dynamics, and the former limits the cell heterogeneity much more than the latter.
Even though the ordered dynamics is slower as consequence of the limits imposed on the transition rates by driver mutations waiting to become effective driver mutations, the ordered and unordered dynamics are comparable up to opportune rescaling of parameters.
Nevertheless, the two dynamics are characterized by completely different types of evolution of the disease after the same kind of drug treatment.
Simulations based on the unordered driver mutations show a slower recovery of the cancer populations and a retardation in the appearance of highly tumorigenic sub--populations, while simulations with ordered driver mutations show a faster restoration of the targeted cells and the anticipation of the appearance of aggressive sub--populations.

The perturbation induced by the drug produces different results on the two dynamics by breaking the symmetries between the heterogeneous cancer cell sub--populations. 
In the examples shown on breast cancer metastasizing in the bone, the restoration induced by the ordered dynamics strongly affects the extravasating cancer cells because, although the cells with mutation in the CD47 gene are sent to apoptosis by the drug, many other cells which present much less effective driver mutations (in this specific case only one effective driver mutation) survive, and some of these surviving cells are mutationally a few steps away from becoming highly tumorigenic. 

Hence, drugs targeting the larger sub--populations of cells just entering the compartment or the sub--populations surviving longer when not treated are less effective in the long time regime because they eliminate the cells with a high number of effective driver mutations. However, they leave cells with an even higher number of driver mutations with a non aggressive phenotype, but which are missing a few mutations to unlock the complete sequence of the right order of mutations and switch to an aggressive tumorigenic phenotype. Furthermore, cells with few effective driver mutations can posses high stem-like potentiality due to acquired driver mutations which are not effective. These cells, in the long time regime, represent a source for many other cancer cells speeding up the restoration of cancer sub--populations.

\section{Conclusion}

We investigated
the role of heterogeneity in cancer cells by embedding a new internal structure driving the development of the disease in the dynamics of the mutation process. Such internal structure is given by the order of occurrence of mutations and the metabolic cell cycle acceleration. The structure introduced in the dynamics stems from a phenomenological derivation of the effects of mutations which results in an increase/decrease of the capability of cancer cells to survive, differentiate, proliferate and metastasise.
On one side, we have considered driver mutations responsible for the increase of cancer cell stemness and their order 
relevant for the development of oncogenic phenotypes involved in the process of metastasization in a secondary site of the cell and its progenies. On the other side, metabolic mutations are related to the production and consumption of ATP and are involved in the acceleration of the mitotic rate.

The order of mutations and acceleration of the dynamics have been considered relevant in some specific types of cancers like melanomas, but their importance and effects have been experimentally difficult to measure. 
To understand their effects on cancer evolution, we introduced an analytical description of the order of driver mutations and the effects of metabolic mutations in terms of operators, and we have included them in the derivation of the master equations for the cancer cell populations.

This model has been applied to the case of breast cancer metastasising in the bone tissue 
for which we simulated the evolution of cancer in presence and absence of the order of driver mutations. 
To highlight the differences between the two types of dynamics, we have compared them before and after they are affected by an external perturbation represented by the action of an anti-tumoral drug. 
The numerical results show the order of mutations introduce a slower dynamics of cancer cells reaching the bone than in simulations with no order of driver mutations. Nevertheless, for realistic drug perturbations of the two dynamics, the recovery of aggressive cancer sub-populations is faster in presence of order of driver mutations than in the case of cancer cells non sensitive to the mutation order.

This model pinpoints how the order of mutations and metabolic mutations in cancer may be responsible for the short time of relapse after drug treatments. Furthermore, this work helps us to understand the difficulties in experimentally detecting the order of mutation in cancer, their role in the dynamics and their response due to the interactions of the system with drugs.

\section*{References}
\bibliographystyle{unsrt}
\bibliography{cancer_models5}

\end{document}